\documentclass[11pt]{article}
\usepackage[T1]{fontenc}
\usepackage[utf8]{inputenc}
\usepackage[margin=24mm]{geometry}
\usepackage{amsmath,amssymb,amsfonts,amsthm}
\numberwithin{equation}{section}
\usepackage{graphicx,tabularx,booktabs,float}
\usepackage[title]{appendix}
\usepackage{tikz}
\usetikzlibrary{arrows.meta,positioning,calc}
\usepackage{xcolor,microtype}
\usepackage[hidelinks,hypertexnames=false]{hyperref}
\begin{document}
\title{Contextual Probability and Operational Minimality in Tug-of-War Decision Making}
\author{Song-Ju Kim\\ SOBIN Institute LLC}
\date{14 October 2026}
\maketitle
\begin{abstract}
What connects minimal decision-making to contextual probability? We address this question through the Tug-of-War (TOW) principle, a resource-coupled mechanism for multi-armed bandit decision-making. Starting from classical probability, we show that when reward-learning functionality and specified information requirements are fixed, minimizing the necessary relaxation of common event semantics can lead to nonclassical probability. We derive the exact minimum semantic repair and identify the conditions under which information efficiency, calibration, and compositional consistency select contextual statistics. We then construct a single quantum Tug-of-War (QTOW) instrument in which the same decision--reward process both improves stationary decision performance and drives the unconditional learned state beyond the Klyachko--Can--Binicio\u{g}lu--Shumovsky (KCBS) noncontextuality bound. A finite diagnostic based on the instrument's recovery operations also yields a corrected contextuality witness. These results connect minimal decision-making, operational economy, and the emergence of contextual probability.
\end{abstract}

\section{Introduction}\label{sec:intro}
Human decisions are often sensitive not only to the available alternatives but also to the circumstances in which they are evaluated. In questionnaire studies, for example, changing the order of two questions can change the proportions of affirmative answers. Systematic question-order effects, together with conjunction and disjunction phenomena, have motivated the use of quantum probability in cognitive modeling \cite{wang2014order,busemeyer2012quantum,pothos2009quantum,BusemeyerTownsend2006JMP}. These results do not establish microscopic quantum activity in the brain. They raise a mathematical question: why can a probability description developed for incompatible measurements be useful for ordinary judgments and decisions?

In the research program known as \emph{quantum cognition}, cognitive states and questions can be represented by states and operations in a Hilbert space. Different questions need not commute, so the order of interrogation may matter. This offers explanations and testable quantitative relations for some behavioral datasets, including the question-order experiments of Wang and colleagues \cite{wang2014order}. Empirical comparisons of Markovian and quantum probabilistic models have also been undertaken for categorization and decision tasks \cite{BusemeyerWangLambert2009}, while a Prisoner's Dilemma experiment has examined a quantum account of the disjunction effect \cite{Tesar2020}. These studies motivate the use of quantum probabilistic models in decision research, but neither by itself constitutes a KCBS test. Yet contextual behavior in its broad sense---including dependence on presentation order or past experience---does not by itself prove Kochen--Specker contextuality. The Kochen--Specker problem, and its operational generalizations, ask more specifically whether common event identities can be maintained across compatible tests \cite{KochenSpecker1967,Spekkens2005}. A classical stochastic model may represent different experimental settings by distinct, context-labelled variables \cite{dzhafarov2015contextuality,DzhafarovKon2018}. Moreover, behavioral measurements may have context-dependent marginals, for which the assumptions of an ideal KCBS experiment must be checked separately. The issue is therefore not whether a classical computer can reproduce a sequence of responses, but which \emph{shared meanings and operational relations} can be retained in a common representation.

This question bears directly on the comparison between natural and artificial intelligence. Natural decision-making takes place through physical dynamics with limited resources, continual feedback and state-dependent responses. Artificial decision makers can implement the same input--output laws using probabilistic programs, stored histories or explicit context labels. Hence the useful contrast is not a categorical assertion that human cognition is quantum whereas artificial intelligence is classical. It is a question about \emph{representation}: when the functions of an adaptive decision maker are specified, which internal distinctions and event identities must be retained, and at what cost? This interpretation is consistent with the broader discussion of quantum-like cognition as a possible connection between natural and artificial intelligence \cite{Khrennikov2026}.

We examine the problem using the Tug-of-War (TOW) principle for multi-armed bandit decision-making. Earlier TOW models introduced resource-coupled learning rules \cite{TOW2009,TOW2010,TOWBio}; the conservation-based TOW principle and an analytical account of its decision-making efficiency were subsequently formulated explicitly by Kim \cite{TOW2015}. The name comes from a physical picture in which a finite resource is allocated between competing options. Following each choice, the external slot machine generates a reward, and the observed result updates the shared internal preference used on later trials. Conservation couples the alternatives, while fluctuations support exploration. TOW is thus a concrete, physics-motivated form of reward-guided adaptation, rather than a theory of all learning or a model of human cognition by itself.

The key question of this work is \emph{what connects this minimal decision process to contextual probability?} We start without quantum postulates. A classical two-state TOW-family realization already provides persistent preferences, reward sensitivity and useful stationary performance. We then hold those native functions fixed and require a specified, calibrated ability to obtain information from the same preference through shared events. For an exclusive five-cycle ($C_5$) \cite{KCBS2008PRL}, each measurement context admits ordinary positive probabilities; nevertheless the five contexts need not admit one common noncontextual assignment. Our first result quantifies the minimum relaxation of shared-event identity and identifies, within a specified information-service family, when the required information crosses the zero-repair boundary. The information constraint and the composition/calibration conditions are declared operational requirements; they are not inferred from bare A/B bandit learning.

A separate question is whether the \emph{learning dynamics itself} can prepare states that display this obstruction, or whether a violating state has simply been chosen for an unrelated measurement. We address it with one recovery-and-reset quantum Tug-of-War (QTOW) instrument. It uses the same A/B readout, external rewards and reward-dependent state update for the entire learning trajectory. The learned \emph{unconditional} state crosses the KCBS bound when interrogated by a fixed five-context probe family. The probe geometry and independently callable recovery controls are part of the explicitly defined extended instrument; ordinary TOW decision dynamics alone neither generates the $C_5$ graph nor forces a violation. This distinction separates a functional selection result obtained with classical probability from a physical quantum realization of a learned-state witness.

Section~\ref{sec:classical-tow} develops the tug-of-war conservation picture and follows a complete choice--reward--update cycle before introducing the minimal stochastic realization of Sec.~\ref{sec:classical-first}. We then derive a minimal common-event repair cost and the information threshold. Section~\ref{sec:classical-selection} discusses the independent role of calibration and two-copy composition; Secs.~\ref{sec:qtow-implementation} and \ref{sec:qtow-reachability} construct and test the same-instrument QTOW learning trajectory. The final discussion connects the results to the quantum-cognition literature and to representational choices in natural and artificial intelligence. Proofs of the central claims are in the Appendices; alternative implementations and auxiliary controls are in the Supplementary Information.

\section{Classical TOW and Reward-Guided Decision-Making}\label{sec:classical-tow}
\subsection{The tug-of-war picture: conservation, choice, and feedback}\label{subsec:tow-principle}
Consider two competing slot machines, or \emph{arms}, $A$ and $B$. At each trial an agent selects one arm, receives a binary reward $r_t\in\{0,1\}$ and updates its estimate of which arm is preferable. The environment supplies the reward probabilistically: $P_A=\Pr(r_t=1\mid k_t=A)$ and $P_B=\Pr(r_t=1\mid k_t=B)$. Throughout, uppercase $A,B$ name the two arms; lowercase $k\in\{A,B\}$ (and $k_t$ at time $t$) denotes a generic or selected arm. Thus $P_k$ denotes $P_A$ or $P_B$ according to the value of $k$; it is not a third probability. These are properties of the external slot machines, not probabilities assigned to the TOW decision itself. The bandit problem is to learn which arm is rewarding by interacting with this initially unknown environment.

The name \emph{tug-of-war} refers to a shared, conserved resource: a gain on one side reduces what remains on the other. In a fluid or deformable-body analogy, let $X_A(t)$ and $X_B(t)$ be the resource volumes allocated to arms $A$ and $B$, while $X_0(t)$ denotes an uncommitted reservoir. The conservation law permits $X_0(t)$ to vary in a more general allocation model, but here we take $X_0(t)=X_0$ fixed for simplicity: the two competing allocations directly exchange a fixed resource $X_A(t)+X_B(t)=X_{\rm tot}-X_0$. Thus
\begin{equation}
 X_{\rm tot}=X_A(t)+X_B(t)+X_0=\mathrm{constant},\qquad
 D_t=X_A(t)-X_B(t).
 \label{eq:tow-resource}
\end{equation}
The \emph{total} resource $X_{\rm tot}$ is distinct from the \emph{difference} $D_t$; since $X_0$ is fixed here, variations of $X_A$ and $X_B$ are equal and opposite. In the simplest two-option comparison, $D_t$ is a noisy physical representation of accumulated evidence, schematically
\begin{equation}
 D_t=Q_A(t)-Q_B(t)+\xi_t,
 \qquad k_t=\begin{cases}A,&D_t>0,\\ B,&D_t<0,\end{cases}
 \label{eq:tow-deterministic-choice}
\end{equation}
where $Q_A,Q_B$ are preference or reward-evidence coordinates in matching units and $\xi_t$ is an instantaneous fluctuation. Here $k_t$ is the selected-arm index, distinct from the fixed arm labels $A$ and $B$. A tie requires a specified convention. \emph{Conditional on the actual fluctuating volumes, the choice is deterministic: the larger side wins.} Repeated choices become probabilistic only when the fluctuations or unknown internal states are averaged over. Volume conservation is the defining intuition; Eq.~\eqref{eq:tow-deterministic-choice} is a schematic comparator, not a claim that every TOW implementation has exactly this noise law.

After an external reward, the evidence variables are modified. One common discounted form is
\begin{equation}
 Q_k(t+1)=\alpha Q_k(t)+U_k(k_t,r_t),\qquad 0\le\alpha\le1,
 \label{eq:tow-discount}
\end{equation}
where $Q_k$ and $U_k$ mean $Q_A,U_A$ for $k=A$, or $Q_B,U_B$ for $k=B$; $U_k$ is the reward-dependent increment prescribed by a particular TOW algorithm and $\alpha$ discounts previous evidence. Resource coupling translates a change of relative evidence into opposite changes of the allocations $X_A$ and $X_B$. The choice, reward sampling and update are therefore three distinct steps, shown in Fig.~\ref{fig:tow-feedback}. The possible time dependence of the true $P_A,P_B$ and the resulting choice of $\alpha$ are relevant to nonstationary bandits, but our explicit stationary-gain theorem below fixes constant $P_A,P_B$.

\begin{figure}[t]
 \centering
 \resizebox{\linewidth}{!}{\begin{tikzpicture}[>=Latex, font=\scriptsize]
  \node[font=\bfseries\footnotesize,anchor=west] at (0.1,3.45) {(a) Conserved resource};
  \draw[thick,rounded corners=1pt] (0.30,.1) rectangle (1.60,2.65);
  \draw[thick,rounded corners=1pt] (1.83,.1) rectangle (3.13,2.65);
  \draw[thick,rounded corners=1pt] (3.36,.1) rectangle (4.66,2.65);
  \fill[blue!22] (.33,.13) rectangle (1.57,2.06);
  \fill[orange!30] (1.86,.13) rectangle (3.10,1.52);
  \fill[gray!25] (3.39,.13) rectangle (4.63,.80);
  \node[font=\bfseries] at (.95,1.0) {$X_A$};
  \node[font=\bfseries] at (2.48,.74) {$X_B$};
  \node[font=\bfseries] at (4.01,.40) {$X_0$};
  \node[anchor=north] at (.95,-.02) {Arm $A$};
  \node[anchor=north] at (2.48,-.02) {Arm $B$};
  \node[anchor=north] at (4.01,-.02) {Fixed reserve};
  \node[align=center] at (2.48,3.02) {$X_A+X_B=X_{\rm tot}-X_0$};
  \draw[<->,very thick,blue!70!black] (1.61,2.29) -- (1.82,2.29);
  \node[font=\bfseries\footnotesize,anchor=west] at (6.2,3.45) {(b) Choice, reward, and feedback};
  \node[draw,rounded corners,fill=blue!8,minimum height=.76cm,text width=1.85cm,align=center,inner sep=3pt] (state) at (7.25,2.22) {Shared preference\\$Q_A,Q_B$ or $h$};
  \node[draw,rounded corners,fill=blue!8,minimum height=.76cm,text width=2.05cm,align=center,inner sep=3pt] (select) at (10.20,2.22) {Noisy comparison\\larger side wins};
  \node[draw,rounded corners,fill=orange!14,minimum height=.76cm,text width=2.15cm,align=center,inner sep=3pt] (slot) at (13.55,2.22) {Selected slot $k_t$\\external reward $r_t$};
  \node[draw,rounded corners,fill=green!12,minimum height=.76cm,text width=3cm,align=center,inner sep=3pt] (update) at (10.25,.45) {Use $k_t,r_t$ to update\\shared preference at $t+1$};
  \draw[->,thick] (state.east) -- (select.west);
  \draw[->,thick] (select.east) -- (slot.west) node[midway,above] {$k_t$};
  \draw[->,thick] (slot.south) |- (update.east) node[pos=.19,right] {$r_t$};
  \draw[->,thick] (update.west) -| (state.south);
  \node[align=center,text width=8.9cm,anchor=north] at (10.35,-.36) {Given the fluctuations, the choice is deterministic.\\Only the slot reward is sampled from $P_A$ or $P_B$.};
\end{tikzpicture}
}
 \caption{\textbf{The classical TOW principle and one bandit trial.} (a) A finite volume is divided among two competing arms and a reserve; moving the resource toward one arm changes the balance without creating volume. The heights are schematic rather than measured physical dimensions. (b) At trial $t$, the current state and a fluctuation produce a \emph{deterministic} comparison of the two sides; the selected external slot machine then generates a \emph{stochastic} reward; this observation updates the shared preference for trial $t+1$. The diagram distinguishes randomness in the physical comparator from randomness in the external reward source.}
 \label{fig:tow-feedback}
\end{figure}

\subsection{From a physical analogy to the probability model used here}\label{subsec:tow-model-scope}
Early TOW learning models preceded physical-device proposals \cite{TOW2009,TOW2010,TOWBio}. The volume-conserving principle was subsequently formulated and analyzed quantitatively in Ref.~\cite{TOW2015}: its shared conservation constraint couples the alternatives, so evidence obtained from one arm can change their relative preference. In parallel with this theoretical development, a quantum-dot near-field excitation-transfer device was proposed numerically in 2013 \cite{Qdot}; later physical implementations and studies included single-photon decision makers \cite{Single}, atomic-switch devices \cite{TOWAS}, and ionic systems \cite{Ion}. These examples show the versatility of the principle without establishing that every implementation has the same internal dynamics.\par
The conservation constraint alone fixes neither the distribution of $\xi_t$ nor the number of states or the reward-update rule. We therefore specify a deliberately small realization with fixed learning capabilities before asking what further \emph{common-event information functions} can be supported consistently. This separates the general TOW principle from the particular model in which our theorems are proved.

Our choice of a binary model is dictated by \emph{operational minimality}, not by a claim that physical TOW is intrinsically discrete. To avoid continuous-valued preference coordinates and to determine the smallest nontrivial internal representation supporting reward-guided adaptation, we retain only a persistent preference sign $h=+1$ or $-1$ in place of the full $Q_A,Q_B$ variables. In a time-homogeneous Markov learner whose persistent internal state carries all learned preference, a single predictive state cannot make a later choice depend on the observed reward; at least two distinct preference states are needed for L1. Our binary construction attains this minimum for the stated native task, without implying a universal compression of continuous TOW. A noisy comparison determines the current choice, and an observed reward partly preserves or resets that preference. Although the resulting choice probabilities are stochastic, they can be understood as averages of deterministic comparisons such as Eq.~\eqref{eq:tow-deterministic-choice}. Section~\ref{subsec:classical-minimal-semantics} defines this model completely and verifies its reward sensitivity, useful stationary adaptation, preference persistence and shared-state operation. It is a deliberately minimal \emph{discrete realization of TOW decision dynamics}, not a universal reduction of continuous TOW nor a model of a mechanically rotating tug-of-war apparatus. The rotations appearing later in QTOW are \emph{state-recovery operations} in a three-mode mathematical representation, not rotations of the classical fluid reservoir.

One further distinction matters throughout: the ordinary choice--reward history is a classical stochastic process even when its state is hidden from an observer. Kochen--Specker contextuality instead concerns the compatibility of \emph{multiple measurement contexts sharing the same events}. The five-cycle diagnostic studied below is a separately specified capability of an extended decision device, rather than five additional slot-machine arms or an automatic consequence of conservation. This is precisely why the classical baseline is informative: it shows what reward learning alone achieves before we impose richer information-retrieval and composition requirements.

\section{Operational Minimality and Classical-First Contextual Probability}
\label{sec:classical-first}

\subsection{Operational minimality depends on the available tasks}
\label{subsec:classical-minimal-semantics}

We fix the capabilities of the decision device before asking which probability structure describes them most economically. The issue is not whether reward learning can be simulated on a classical computer, but whether one \emph{common event semantics} suffices for a specified family of choices, updates and information-retrieval tasks. We begin without assuming any Hilbert-space state, quantum measurement or Born probability.

Let $\mathcal S$ be the set of allowed preparations and $\mathcal{O}$ the set of admissible experimental continuations. Two preparations $s$ and $s'$ are operationally equivalent, written $s\sim_{\mathcal{O}}s'$, if they assign the same probabilities to all observable outcome strings generated by continuations in $\mathcal{O}$. The operationally minimal state space is the collection of equivalence classes $\mathcal{S}/\!\sim_{\mathcal{O}}$. Enlarging $\mathcal{O}$ may distinguish preparations that previously had the same operational description. This is a statement about the economy of an operational representation; it is not an assumption that underlying classical ontic distributions must be identical whenever two preparations are operationally equivalent.

For the native adaptive two-arm bandit-decision service, we fix four modest functional requirements. These characterize the particular reward-guided decision task, not learning in general; later requirements on information retrieval will be stated separately:
\begin{description}
\item[\textbf{L1 (reward sensitivity).}] For at least one reachable preference and selected action, different observed reward outcomes lead to different probabilities of a subsequent choice.
\item[\textbf{L2 (useful adaptation).}] For unequal Bernoulli reward probabilities $P_A\ne P_B$, the stationary expected reward exceeds that of uninformed $50/50$ action selection.
\item[\textbf{L3 (preference persistence).}] An informative choice/readout does not erase every distinction between previously learned preferences: distinct prior preferences can remain distinguishable after the readout and its associated update.
\item[\textbf{L4 (shared state semantics).}] Decision probabilities and feedback updates refer to the same persistent internal preference state; they do not rely on separate, independently maintained preference states for different operations.
\end{description}
These conditions require neither asymptotically optimal learning nor projective measurements. The following fully classical model satisfies them. We deliberately ask for a \emph{minimal finite-state realization}: retaining continuous $Q_A,Q_B$ is unnecessary to prove the basic existence and lower-complexity claims. Within a time-homogeneous Markov model whose persistent preference is its only learned-state resource, one internal state cannot encode a reward-dependent change of the next choice, whereas the two states constructed below suffice. Importantly, L4 concerns the native decision--learning service and does not, by itself, supply five contextual diagnostic events.

Accordingly, the two-state representation retains only the \emph{sign} of the allocated preference: $h=+1$ means a preference for $A$, while $h=-1$ means a preference for $B$. This latent variable is persistent across trials; it is not itself the selected action, the physical volume difference, or an extra random reward. For the arm index $k\in\{A,B\}$ defined in Sec.~\ref{subsec:tow-principle}, introduce the \emph{arm-sign function} $s_k$ by $s_A=+1$ and $s_B=-1$. Thus $s_k$ is the indexed notation for these two values, not a separate state variable; the same function $s_k$ appears in the choice and update rules. The distinct dimensionless parameter $\chi$, with $0<\chi<1$, measures the readout contrast: it controls how strongly the choice follows the stored preference. The symbol $\chi$ is neither an arm label nor an external reward probability $P_A$ or $P_B$.

One can generate the choice by an entirely deterministic comparison once a new fluctuation $\varepsilon_t\sim\mathrm{Unif}[-1,1]$ is realized: select $A$ when $\chi h+\varepsilon_t>0$, and $B$ otherwise. Integrating over this fluctuation gives the explicit \emph{conditional} choice probability
\begin{equation}
 \Pr(k\mid h)=\frac{1+s_k \chi h}{2}.
 \label{eq:cf-native-choice}
\end{equation}
For this binary abstraction, the normalized allocated amounts can be written $x_A=(1+h)/2$ and $x_B=(1-h)/2$, so $x_A+x_B=1$ for either latent state and after every update. These $x_k$ are \emph{logical allocation coordinates}, not literal measurements of the fluid heights in Fig.~\ref{fig:tow-feedback}. Their conserved sum is the discrete analogue of the original volume-coupling principle.

The outside world then returns a binary reward $r\in\{0,1\}$, with $\Pr(r=1\mid k)=P_k$, where $P_k$ is precisely $P_A$ or $P_B$ for the selected arm (Sec.~\ref{subsec:tow-principle}). With probability $1-\gamma$ the device retains its old sign $h$; with probability $\gamma$ it resets to $s_k(2r-1)$. Thus a reward ($r=1$) reinforces the just-chosen arm, while a failure ($r=0$) favors the alternative. The parameter $\gamma\in(0,1)$ is the probability of applying this feedback reset, so it governs persistence in this \emph{particular stochastic realization}; it is not the same parameter as the discount $\alpha$ of the continuous $Q$ update in Eq.~\eqref{eq:tow-discount}. Figure~\ref{fig:tow-feedback} also makes clear why the choices and rewards have different sources of randomness. The complete probability of one trial is the product of (i) choosing an arm from the noisy readout, (ii) sampling that arm's external reward, and (iii) keeping or resetting the hidden preference. This yields the ordinary nonnegative, outcome-labelled Markov kernels
\begin{equation}
 [T_{k,r}]_{h,h'} =
 \frac{1+s_k \chi h}{2}\,
 P_k^{\,r}(1-P_k)^{1-r}
 \Big[(1-\gamma)\mathbf{1}_{h'=h}
 +\gamma\mathbf{1}_{h'=s_k(2r-1)}\Big].
 \label{eq:cf-native-kernel}
\end{equation}
The first factor in Eq.~\eqref{eq:cf-native-kernel} is the choice probability, the second is the reward likelihood, and the bracketed third factor is the conditional update from the old latent sign $h$ to the next sign $h'$. The kernel is defined for every selected arm $k$ and observed reward $r$; the summation over all $(k,r,h')$ is one for each old $h$. If $\pi_0$ is the initial row distribution over $h$ and $\mathbf 1$ the two-component all-ones terminal column, every finite action--reward history consequently has a conventional Kolmogorov probability:
\begin{equation*}
 \Pr(k_1,r_1,\ldots,k_n,r_n)=
 \pi_0 T_{k_1,r_1}\cdots T_{k_n,r_n}\mathbf 1.
\end{equation*}

Although the internal sign $h$ of a single trial is either $+1$ or $-1$, the model also has an ensemble statistic; it does not add a continuously valued memory register to the individual two-state learner. Its \emph{mean polarization} $m_t=\Pr(h_t=+1)-\Pr(h_t=-1)=\mathbb E[h_t]$ lies in $[-1,1]$; it records the population-averaged balance of preferences and will be the bridge to the QTOW density operator below. Averaging the trial kernel over choices and rewards gives
\begin{equation}
 m_{t+1}
 =\left[1-\gamma+\gamma \chi(P_A+P_B-1)\right]m_t
  +\gamma(P_A-P_B).
 \label{eq:cf-m-recurrence}
\end{equation}
For constant Bernoulli parameters and a stable update, the stationary mean $m_*$ is the fixed point of this recurrence. The expected probability of choosing $A$ is $(1+\chi m_t)/2$; substituting it into the expected reward and comparing with uniform random choice gives the stationary improvement
\begin{equation}
 m_*=\frac{P_A-P_B}{1-\chi(P_A+P_B-1)},\qquad
 G_*=\frac{\chi(P_A-P_B)^2}{2[1-\chi(P_A+P_B-1)]}>0
 \quad(P_A\ne P_B).
 \label{eq:cf-native-learning-gain}
\end{equation}
When both reward outcomes have nonzero probability for a selected arm, changing the observed reward from $0$ to $1$ changes the next-trial probability of choosing $A$ by $\chi\gamma$ in magnitude, establishing L1 for the admissible environments. Equation~\eqref{eq:cf-native-learning-gain} establishes L2. For L3, the conditioned mean preference after observing a choice $k$ and reward $r$ is
\begin{equation}
 m'_{k,r}=(1-\gamma)\frac{m+s_k \chi}{1+s_k \chi m}+\gamma s_k(2r-1),
 \qquad \frac{\partial m'_{k,r}}{\partial m}
 =\frac{(1-\gamma)(1-\chi^2)}{(1+s_k \chi m)^2}>0,
 \label{eq:cf-conditioned-preference-persistence}
\end{equation}
so distinct prior preferences remain distinguishable within the supported range. L4 follows because the same latent variable $h$ determines both choice probabilities and the feedback kernel. For example, with $P_A=0.8$, $P_B=0.2$, $\chi=0.6$ and $\gamma=0.1$, the fixed point is $m_*=0.6$, the probability of choosing the better arm is $0.68$, and expected reward is $0.608>0.5$. Thus the countermodel is a genuinely useful, though not optimal, bandit decision maker. Its purpose is to show that exactly two persistent preference states \emph{suffice} for L1--L4 and, within the specified time-homogeneous shared-state Markov class, are the minimum needed for reward-sensitive adaptation. This is not a claim that all continuous TOW machines can be reduced to two states or that all their learning capabilities are retained. The same kernel governs native action--reward histories in the logical code of the recovery-and-reset QTOW instrument introduced in Sec.~\ref{sec:qtow-implementation}, with a different numerical contrast $\chi$ chosen there to permit its physical recovery construction. Holding these native functions fixed allows us to ask a sharper question: if an information-retrieval service on the same persistent preference is also prescribed, what is the least modification of a common noncontextual event semantics needed to preserve the combined functions?

\subsection{Shared-event semantics and the classical gluing problem}
\label{subsec:classical-gluing}

The native learner can choose an arm and update its preference without needing to explain that preference in any other way. Yet a learned state may also be used to answer questions. For example, an observer may inquire about a particular feature of the stored preference while pairing that question with different, compatible questions. If the answer is intended to describe the \emph{same feature} in both arrangements, its meaning should not depend on which other question accompanies it. This is a demand on the reusable description of the internal state, beyond the ability to generate a correct sequence of A/B choices and rewards.

We make that demand precise through a diagnostic service with binary Yes/No events. Two events are \emph{compatible} if they can be tested in a common measurement arrangement, or \emph{context}; they are \emph{exclusive} if both cannot return Yes in one such measurement. Reusing one event in two contexts allows us to ask whether locally valid probability tables can be read as statements about one and the same event. A chain of contexts may conceal an inconsistency that becomes apparent on closing a loop. The five-cycle $C_5$ is the smallest nontrivial odd exclusivity cycle of this kind: it makes the difference between separately consistent pairs and a single consistent assignment to all the events particularly transparent \cite{KCBS2008PRL}.

We therefore specify five shared events as an additional information-retrieval capability of the same decision-making architecture. This extension describes \emph{what the device is asked to report} about its learned state; it is not a consequence of the original bandit update. Let its Yes/No events be $e_0,\ldots,e_4$. The available joint contexts form the five-cycle graph $C_5$:
$c_i=(e_i,e_{i+1})$, where indices are modulo five, and neighboring Yes events
are exclusive (Fig.~\ref{fig:cf-c5-pentagon}). The edges represent the joint
measurement contexts, whereas the vertices represent shared event identities:
$e_i$ participates both in $c_{i-1}$ and in $c_i$. The graph is the
\emph{specified diagnostic task}; it is not generated by the native TOW
choice--reward dynamics. The labels $e_i$ are shared: each event appears in two neighboring
contexts with the same marginal Yes probability $p_i$.
\begin{figure}[H]
\centering
\begin{tikzpicture}[font=\footnotesize,
  ev/.style={circle,draw=black!75,fill=white,minimum size=6.7mm,inner sep=0.5pt}]
  \foreach \i in {0,...,4}{
    \coordinate (v\i) at ({1.45*cos(90-72*\i)},{1.45*sin(90-72*\i)});
    \node[ev] (e\i) at (v\i) {$e_{\i}$};
  }
  \foreach \i/\j in {0/1,1/2,2/3,3/4,4/0}{
    \draw[line width=1.05pt,black!70] (e\i)--(e\j);
  }
  \node[font=\scriptsize,anchor=west]  at (1.05,1.07) {$c_0$};
  \node[font=\scriptsize,anchor=west]  at (1.47,-0.45) {$c_1$};
  \node[font=\scriptsize,anchor=north] at (0,-1.88) {$c_2$};
  \node[font=\scriptsize,anchor=east]  at (-1.47,-0.45) {$c_3$};
  \node[font=\scriptsize,anchor=east]  at (-1.05,1.07) {$c_4$};
\end{tikzpicture}

\caption{The five-cycle exclusivity graph $C_5$ used in the diagnostic service. Each vertex denotes a shared Yes/No event $e_i$; each edge denotes a jointly measured exclusive pair $c_i=(e_i,e_{i+1})$ (indices modulo five). The five contexts use the same event labels, rather than five unrelated pairs of outcomes.}
\label{fig:cf-c5-pentagon}
\end{figure}
We assume
\begin{equation}
 p_i\ge 0,\qquad p_i+p_{i+1}\le1 .
 \label{eq:cf-c5-local-conditions}
\end{equation}
Each context separately admits an \emph{ordinary classical} probability table,
\begin{equation}
 \begin{array}{c|cccc}
 (e_i,e_{i+1}) &(1,0)&(0,1)&(0,0)&(1,1)\\ \hline
 \Pr(\cdot\mid c_i)&p_i&p_{i+1}&1-p_i-p_{i+1}&0 .
 \end{array}
 \label{eq:cf-c5-tables}
\end{equation}
For example, in the context $(e_0,e_1)$ the outcomes Yes--No, No--Yes and No--No have probabilities $p_0,p_1$ and $1-p_0-p_1$, respectively. Exactly the same event $e_1$ also appears in context $(e_1,e_2)$, where its Yes probability must agree with the previous one in the ideal consistently connected case. The five-cycle is a \emph{chosen diagnostic structure}, not five reward machines or a graph forced by the TOW conservation equation. Local nonnegativity of these ordinary tables is weaker than the existence of a single joint assignment to all five event labels.

No Hilbert space or Born rule is involved in these tables. In the global-section approach to contextuality
\cite{AbramskyBrandenburger2011}, we ask whether their common event labels can \emph{also} be represented by five jointly
defined, context-independent binary variables $X_0,\ldots,X_4$ obeying
$X_iX_{i+1}=0$ on every run.

\noindent\textbf{Proposition 1 (gluing boundary for the exclusive pentagon).}
Under Eqs.~\eqref{eq:cf-c5-local-conditions}--\eqref{eq:cf-c5-tables},
a positive global distribution over the five context-independent binary
variables exists if and only if
\begin{equation}
 K:=\sum_{i=0}^{4}p_i\le2.
 \label{eq:cf-gluing}
\end{equation}
\emph{Proof sketch.} Any allowed joint binary assignment corresponds to an
independent set of the pentagon and contains at most two Yes values, so
$K\le2$ is necessary. Together with nonnegativity and the neighboring-pair
inequalities, $K\le2$ also characterizes the stable-set polytope of $C_5$,
establishing sufficiency. A constructive description appears in Appendix~\ref{app:classical-first}.
\hfill$\square$

This is a \emph{restricted} necessary-and-sufficient result for the ideal,
consistently connected, exclusive five-event marginal scenario; it does not
assert that satisfying the KCBS inequality proves noncontextuality for arbitrary
sequential measurements, preparations, instruments, or behavioral datasets.
Moreover, failure of a common outcome-value assignment does not preclude a
context-aware classical sampler implementing Eq.~\eqref{eq:cf-c5-tables}.

\subsection{How much common-event semantics must be relaxed?}
\label{subsec:classical-semantic-defect}

For fixed information-service probabilities $p$, the least departure from perfect common-event identity is an optimization problem. To measure it, split each shared event into its two context-labelled occurrences $X_i^L,X_i^R$.
For any joint \emph{coupling} $\Gamma$ of these ten binary variables that
reproduces all five local exclusive pair distributions, define the
\emph{event-identity mismatch}
\begin{equation}
 D_{\rm sem}(\Gamma)=\sum_{i=0}^4
  \Pr_\Gamma(X_i^L\ne X_i^R),\qquad
 D_{\rm sem}^{\min}(p)=\min_\Gamma D_{\rm sem}(\Gamma).
 \label{eq:cf-semantic-defect}
\end{equation}

\noindent\textbf{Proposition 2 (exact minimal semantic-repair cost).}
For every consistently connected exclusive $C_5$ empirical model satisfying
Eq.~\eqref{eq:cf-c5-local-conditions},
\begin{equation}
 \boxed{\displaystyle
 D_{\rm sem}^{\min}(p)=2\,[K-2]_+,\qquad [x]_+=\max\{0,x\}.
 }
 \label{eq:cf-semantic-cost}
\end{equation}
\emph{Proof sketch.} Put $Z_i=\min(X_i^L,X_i^R)$. The local exclusivity
constraints imply $Z_iZ_{i+1}=0$, and therefore $\sum_i Z_i\le2$
for each coupled sample. The identity
$(X_i^L+X_i^R)/2=Z_i+|X_i^L-X_i^R|/2$
gives $D_{\rm sem}\ge2(K-2)$. Nonnegativity gives the bound at $K\le2$;
the bound is sharp by an explicit coupling that splits only one event,
obtained from the stable-set/half-vector decomposition of the pentagon.
\hfill$\square$

Equation~\eqref{eq:cf-semantic-cost} solves the semantic-minimality problem for any \emph{fixed} exclusive and consistently connected five-event table: zero repair is possible exactly when $K\le2$; above that boundary the smallest unavoidable event-identity disagreement is strictly positive. This is a probability of mismatch in a coupling, not physical memory in bits, ontic-state cardinality or unrestricted classical computation cost. The theorem does not choose the event probabilities by itself. In particular, one can select a perfectly valid $C_5$ table with $K\le2$ while keeping the entire native bandit learner unchanged; the mere availability of a pentagon does \emph{not} cause contextuality. What remains is to determine whether independently specified information requirements place the resulting $p_i$ beyond the zero-repair boundary. This is the distinction between a witness scenario that \emph{tests} contextuality and a functional condition that \emph{selects} statistics exhibiting it.

\paragraph{A second, information-theoretic semantic cost.}
The mismatch in Eq.~\eqref{eq:cf-semantic-cost} is not the only way to quantify the same failure of a common-event description. In a semantics-preserving completion, the context $C$ is sampled independently of a shared hidden variable $\Lambda$, and a label $M$ is used to select a stable five-bit valuation for the requested pair. For the specified class with uniformly sampled contexts and $K>2$,
\begin{equation}
 J_{\rm sem}:=\inf I(C;M\mid\Lambda)\ \ge\
 D_{\rm Ber}\!\left(\frac{2K}{5}\,\middle\|\,\frac45\right).
 \label{eq:new-semantic-info-bound}
\end{equation}
The bound is tight for symmetric $C_5$ tables. The information measure is defined using the base-2 binary relative entropy $D_{\rm Ber}(u\|v)=u\log_2(u/v)+(1-u)\log_2[(1-u)/(1-v)]$. Its proof and the precise competitor class appear in Supplementary Information (Sec. S2). It measures the information needed by that class to implement the specified shared-event interpretation.

\paragraph{Connection to a reward-driven classical learner.}
The two-state learner in Eq.~\eqref{eq:cf-native-kernel} can be paired with an independently specified five-context diagnostic. In the constructed example, the learned mean preference $m$ controls $K(m)=(35+5m)/18$: the diagnostic crosses $K=2$ when $m>1/5$, whereas a comparison diagnostic with the \emph{same learner and sensitivity} remains noncontextual. In particular, a learned $m=0.6$ gives $K=19/9$, $D_{\rm sem}^{\min}=2/9$, and the conditional information bound $J_{\rm sem}\ge0.009469$ bits. The diagnostic response probabilities and contexts are \emph{additional choices}; the construction does not mean that TOW learning itself has a non-Kolmogorov probability law. The complete probability tables, matched control, and sequential classical realization remain in Supplementary Information (Sec. S1). Thus the learned preference controls the degree of diagnostic nonextendibility in this constructed service; it does not convert the underlying bandit update into a non-Kolmogorov process.

\subsection{Fixed information requirements and minimal semantic repair}
\label{subsec:classical-selection-functional}
The gluing result tells us how to evaluate an already specified table of shared-event probabilities, but it does not explain which table a useful diagnostic would produce. In particular, a device could avoid semantic repair by returning a sufficiently uninformative answer about its stored preference. To exclude that trivial solution, we now prescribe how informative the diagnostic must be. The learning dynamics remains fixed; the design question is how little common-event consistency needs to be relaxed while meeting a specified information requirement and the stated calibration and composition conditions. This is an optimization over an \emph{additional information service}, formulated entirely with classical probabilities. No Hilbert-space representation is required for its derivation.

Keep the same native two-state decision--reward kernel of Eq.~\eqref{eq:cf-native-kernel}. Suppose the device also supports a nondisturbing five-context three-outcome diagnostic whose shared-event Yes response is
\begin{equation}
 r_i(h)=\frac13+\epsilon d_i+\frac\kappa2 h,\qquad
 d=(-2,-1,0,1,2),\quad\epsilon=\frac1{2000},\quad h=\pm1.
 \label{eq:functional-calibration}
\end{equation}
Here $1/3$ is the chosen neutral reference for a three-outcome measurement: before a preference has developed, its two individual Yes events have probabilities close to one third, as does the remaining outcome. This is a calibration choice, not a value dictated by L1--L4. The small offsets $\epsilon d_i$ distinguish the five event labels while preserving their zero-sum baseline, and $\kappa\ge0$ controls how strongly the diagnostic reports the stored preference. Each context returns the two exclusive neighboring Yes outcomes or neither, exactly as in Eq.~\eqref{eq:cf-c5-tables}, replacing $p_i$ by $r_i(h)$. We require two-copy global exclusivity (GE2) also for accessible independent preparations. Within this family it gives the sharp admissible contrast bound
\begin{equation}
 \kappa\le\kappa_{\rm GE}:=2\left(\sqrt{\frac15-\epsilon^2}-\frac13\right)
 =0.2277599653\ldots.
 \label{eq:functional-GE-bound}
\end{equation}
The readout-strength parameter $\kappa$ has an operational meaning: it sets how differently the diagnostic responds to $h=+1$ and $h=-1$. A larger $\kappa$ can distinguish learned preferences more effectively, but it must remain within nonnegative, calibrated probabilities and GE2. For a neutrally mixed preference $\Pr(h=\pm1)=1/2$ and an independently uniformly chosen context $C$, define $I_0(\kappa):=I(H;O\mid C)$, where $O$ is the three-valued diagnostic result. This is information \emph{about the stored preference} made accessible by the stipulated service, not the information about external reward probabilities acquired in native bandit learning. Every weaker contrast channel is a strict stochastic garbling of a stronger one, so $I_0(\kappa)$ is strictly increasing on the admissible interval. In particular, maximizing this information selects $\kappa=\kappa_{\rm GE}$. The learned sum is
\begin{equation}
 K(m)=\frac53+\frac{5\kappa_{\rm GE}}2m,
 \quad K(m)>2\ \Longleftrightarrow\ m>\frac{2}{15\kappa_{\rm GE}}=0.58541163\ldots.
 \label{eq:functional-crossing}
\end{equation}
For $P_A=0.8$, $P_B=0.2$, $\gamma=0.1$ and $m_0=0$, the ordinary learner has $m_t=0.6(1-0.9^t)$, whatever fixed contrast $\chi$ is chosen when $P_A+P_B=1$. Thus the efficient diagnostic selects $K(m_*)=2.008306615\ldots>2$ and first exceeds two after trial $36$ (Appendix~\ref{app:functional-selection}).

\paragraph{The fixed-information minimal-repair problem.}
For any fixed learned polarization $m>0$, let $I_{\rm req}$ be the information that the calibrated service must provide about the stored preference, evaluated at the neutral input prior used to define $I_0$. Holding the native learning kernel, mean preference $m$, neutral calibration and GE2 admissibility fixed, define
\begin{equation}
 \mathcal D_m^*(I_{\rm req})=
 \min_{\substack{0\le\kappa\le\kappa_{\rm GE}\\I_0(\kappa)\ge I_{\rm req}}}
 D_{\rm sem}^{\min}\bigl(p(\kappa,m)\bigr).
 \label{eq:cf-fixed-info-repair-problem}
\end{equation}
Since $I_0$ increases strictly with $\kappa$ and
$D_{\rm sem}^{\min}=[5m\kappa-2/3]_+$ is nondecreasing, the minimum is attained at
$\kappa_{\rm req}=I_0^{-1}(I_{\rm req})$. Hence, for
$0\le I_{\rm req}\le I_0(\kappa_{\rm GE})$,
\begin{equation}
 \boxed{\displaystyle
 \mathcal D_m^*(I_{\rm req})=
 \left[5m I_0^{-1}(I_{\rm req})-\frac23\right]_+.}
 \label{eq:cf-fixed-info-repair-result}
\end{equation}
For $m>0$, the zero-repair boundary is $\kappa_{\rm crit}=2/(15m)$ when this contrast lies within the admissible interval. In the specified stationary learner with $m_*=0.6$, its neutral-prior information threshold is $I_0(\kappa_{\rm crit})=0.1711306386\ldots$ bits, whereas the maximal allowed information is $I_0(\kappa_{\rm GE})=0.1804873033\ldots$ bits. Consequently, maintaining the fixed native reward-learning functions and demanding more information than this threshold forces $\mathcal D_{m_*}^*>0$ \emph{within the calibrated GE2-admissible family}. At the information optimum the result agrees with Eq.~\eqref{eq:functional-crossing}: $K(m_*)=2.008306615\ldots$ and $D_{\rm sem}^{\min}=0.016613229\ldots$.

The information-maximization and fixed-information minimal-repair formulations thus describe the same achievable information--semantic-cost boundary of this family; they are not a general duality between physical memory and learning performance. Outside the stipulated calibration, event and GE2 requirements, a noncontextual service with the same native learner exists, and unrestricted diagnostic loss can eliminate the violation. These controls, the strictly increasing information function and the GE2 calculation are detailed in Appendix~\ref{app:functional-selection}. Nothing here asserts that every selected local table must be realized by the qutrit implementation of Sec.~\ref{sec:qtow-implementation}.

\section{Conditions for a Quantum-Compatible KCBS Point}\label{sec:classical-selection}
The preceding analysis begins with a required amount of information about the learned preference and asks how much shared-event meaning can be preserved. A related, but different, issue is how a family of diagnostic probabilities should behave when experiments are combined and its responses are calibrated. Local consistency of each $C_5$ pair does not settle this issue: the same readout may be used on separately prepared copies, and the probabilities of the resulting product events must then remain coherent. We first express that compositional requirement as two-copy global exclusivity (GE2). We then examine an explicitly defined calibration objective to determine which probability vector lies on the allowed boundary. Figure~\ref{fig:cf-ge2-calibration-map} summarizes these two additional conditions for the diagnostic of Fig.~\ref{fig:cf-c5-pentagon}.
\begin{figure}[t]
\centering
\resizebox{.90\linewidth}{!}{
\begin{tikzpicture}[font=\footnotesize,>=Latex,
  lab/.style={font=\footnotesize,align=center},
  sub/.style={font=\scriptsize,align=center,text=black!68},
  arrow/.style={-{Latex[length=2mm]},line width=.7pt,draw=black!55}]
\node[lab] at (2.1,3.65) {Two independent $C_5$ copies};
\foreach \i in {0,...,4}{
  \foreach \j in {0,...,4}{
    \fill[black!22] ({1.20+.45*\i},{2.85-.45*\j}) circle (0.091);
  }
}
\foreach \i/\j in {0/0,1/2,2/4,3/1,4/3}{
  \filldraw[fill=teal!68!black,draw=teal!68!black] ({1.20+.45*\i},{2.85-.45*\j}) circle (0.120);
}
\node[lab] at (2.1,.40) {$K^2\leq5$};
\node[sub] at (2.1,.04) {Global exclusivity (GE2)};
\draw[arrow] (3.85,1.94)--(4.65,1.94);
\node[sub] at (4.24,2.34) {calibrate};
\node[lab] at (7.4,3.65) {Efficient calibration};
\coordinate (O) at (5.85,.70);
\draw[->,black!75] (O)--(9.45,.70) node[right,font=\footnotesize] {$\vartheta$};
\draw[->,black!75] (O)--(5.85,3.14) node[left,font=\footnotesize] {$\tau$};
\fill[teal!9] (O)--(8.14,.70) arc[start angle=0,end angle=90,radius=2.29]--cycle;
\draw[teal!75!black,line width=1.45pt] (8.14,.70) arc[start angle=0,end angle=90,radius=2.29];
\draw[dashed,black!36] (5.85,2.99)--(7.78,2.99);
\fill[teal!75!black] (5.85,2.99) circle (0.095);
\node[anchor=west,font=\scriptsize,align=left] at (6.08,2.70) {$\tau(0)=1/\sqrt5$};
\node[anchor=west,font=\scriptsize,align=left] at (6.57,1.74) {$\tau^2+\vartheta^2\leq 1/5$};
\node[sub] at (7.40,.23) {Efficiency selects the boundary};
\end{tikzpicture}
}
\caption{Additional operational conditions imposed on the $C_5$ diagnostic of Fig.~\ref{fig:cf-c5-pentagon}. Two independently prepared copies supply product events (left); the highlighted exclusive clique is part of the GE2 constraint $K^2\le5$. An independently specified efficient calibration has a quadratic probability budget (right; schematic, not to scale). Neither condition is entailed by native TOW reward learning.}
\label{fig:cf-ge2-calibration-map}
\end{figure} Local exclusivity alone permits $K\le5/2$, whereas a global noncontextual assignment requires $K\le2$.

Imagine interrogating two independently prepared copies in parallel. ``Yes for event $i$ on the first copy \emph{and} Yes for event $j$ on the second'' is now a product event, with probability $p_ip_j$ under the stated independence assumption. Two product events are exclusive whenever at least one of their component events cannot occur together. We ask that every such exclusive set still obey the ordinary unit probability budget. This is \emph{two-copy global exclusivity} (GE2): a consistency condition on the composed diagnostic service, stronger than separately normalizing each one-copy context. Because the 25 product events can be partitioned into five mutually exclusive cliques, GE2 gives
\begin{equation}
 K^2=\Bigl(\sum_{i=0}^{4}p_i\Bigr)^2\le5,
 \qquad K\le\sqrt5.
 \label{eq:cf-ge2-k}
\end{equation}
This is the known global-exclusivity bound \cite{Cabello2013GE,CabelloSeveriniWinter2014}. It arises from how independent tests can be composed, not from the reward probabilities of the two slot machines. In particular, the one-copy conditions allow $K$ as large as $5/2$, while the two-copy requirement limits it to $\sqrt5$. The composed experiment is therefore informative about which probability descriptions remain admissible when the same service is reused.

GE2 alone, however, limits the sum $K$ without determining the shape of the five-component probability vector. To study selection rather than mere admissibility, we consider a family of calibrations in which the relative Yes rates can be varied in a controlled way. Specifically, let a small adjustable asymmetry be denoted by $\vartheta$, and set
\begin{equation}
 r_i(\vartheta)=\tau(\vartheta)+\vartheta v_i,
 \qquad v=(2,-\tfrac12,-\tfrac12,-\tfrac12,-\tfrac12),
 \quad 0\le\vartheta\le10^{-2}.
 \label{eq:cf-cal-family}
\end{equation}
The common component $\tau$ sets the overall Yes rate, whereas $\vartheta v_i$ changes how that rate is distributed among the five labelled events. We choose the zero-sum direction $v$ so that varying $\vartheta$ redistributes weight without changing the total $K=5\tau$; its squared components sum to five, making the compositional constraint especially transparent. This particular direction is a convenient \emph{specified calibration family}, not a unique physical response prescribed for all decision makers. Applying GE2 to the five product events $(i,2i)$ of two copies of this preparation yields
\begin{equation}
 5\tau(\vartheta)^2+5\vartheta^2\le1.
 \label{eq:cf-cal-envelope}
\end{equation}
At fixed $\vartheta$, a larger $\tau$ makes the calibrated service return Yes more often without changing the chosen asymmetry pattern. We call the calibration \emph{efficient} when it maximizes precisely this aggregate Yes rate, $K=5\tau$, within the GE2 budget. The resulting boundary response is
\begin{equation}
 \tau_{\mathrm{eff}}(\vartheta)=\sqrt{\tfrac15-\vartheta^2},
 \qquad r_i^{\mathrm{eff}}(\vartheta)=\tau_{\mathrm{eff}}(\vartheta)+\vartheta v_i.
 \label{eq:cf-efficient-cal-family}
\end{equation}
Efficiency here has a narrow operational definition: maximize the aggregate Yes rate \emph{at a fixed, deliberately controlled asymmetry}. It is not a general principle that cognitive agents maximize this quantity, nor is it the same optimization as the mutual information $I_0(\kappa)$ about the internal preference in Sec.~\ref{subsec:classical-selection-functional}. The latter fixes an information-retrieval requirement; the present calibration explores the largest response compatible with composition. Both optimization problems start from ordinary classical probabilities, but they constrain different aspects of the diagnostic service. The family in Eq.~\eqref{eq:cf-efficient-cal-family} satisfies the required local and pairwise GE2 conditions on the stated interval, as shown in Appendix~\ref{app:calibration-details}.

A further consequence is \emph{conditional rigidity}: if the accessible preparation service includes arbitrarily small positive settings of this efficient family and all accessible preparation pairs satisfy mixed-copy GE2, then any accessible $p$ attaining $K=\sqrt5$ must obey
\begin{equation}
 p_0=p_1=\cdots=p_4=\frac1{\sqrt5}.
 \label{eq:cf-rigidity}
\end{equation}
The calibration calculation and finite-setting bound are given in Appendix~\ref{app:calibration-details}. Thus GE2 supplies a compositional probability bound, and the controlled efficient family makes the maximizer rigid within the stipulated service. The resulting uniform vector is compatible with the quantum KCBS maximum; the argument itself uses neither a Hilbert-space model nor a physical learning update.

These classical-first results lead to a separate question. We now know how to \emph{test} whether shared event meanings can be maintained and how specified information, calibration and composition requirements constrain their probability tables. We have not yet shown that a working decision-making mechanism can \emph{learn a state} whose statistics cross the noncontextual boundary under an unchanged diagnostic. Sections~\ref{sec:qtow-implementation} and \ref{sec:qtow-reachability} address this dynamical realization problem with a single physical QTOW instrument. Quantum states and operations enter at this implementation stage, while the operational meaning and mathematical scope of the preceding classical bounds remain unchanged.

\section{A Single QTOW Instrument for Reward-Guided Decisions}
\label{sec:qtow-implementation}

The classical analysis identifies when specified information demands prevent a single noncontextual meaning for shared events. It does not tell us whether a decision maker can \emph{learn its way} into a state displaying such statistics. This is the purpose of the quantum Tug-of-War (QTOW) construction: instead of preparing a convenient contextual state independently of the task, we let the same agent repeatedly choose an arm, observe the external reward, and update its retained preference. Only after those learning trials do we test the resulting state with a separate, fixed set of compatible diagnostic contexts.

The physical task is to preserve a useful preference despite repeatedly interrogating it. A binary detector must distinguish arms $A$ and $B$; the act of reading it can disturb the state that should survive to the next trial. We therefore combine each readout with a fixed recovery operation, followed by a reward-conditioned update. Figure~\ref{fig:qtow-operation-cycle} separates this native learning cycle from the optional five-context diagnostic. This distinction also keeps QTOW's three quantum modes separate from the conserved volumes in the classical TOW picture: they are alternative realizations of resource-coupled decision principles, not the same physical variables. The numerical contrast in the present QTOW example is $\chi=5\sqrt7/16$, rather than $\chi=3/5$ in the classical information-selection example; their stationary rewards must not be identified.

\begin{figure}[H]
\centering
\resizebox{\linewidth}{!}{\begin{tikzpicture}[
  x=1cm,y=1cm,font=\sffamily\small,
  box/.style={draw=black!75,line width=.65pt,rounded corners=2pt,
              fill=black!3,align=center,minimum height=1.15cm,inner sep=5pt},
  flow/.style={-{Latex[length=2.1mm,width=1.5mm]},line width=.9pt},
  probe/.style={flow,dashed,black!70}
]
\node[box,text width=2.38cm] (state) at (1.36,0)
 {\textbf{Internal state}\\[2pt]$\rho_t$\\[-1pt]{\footnotesize logical preference}};
\node[box,text width=2.52cm] (detect) at (4.87,0)
 {\textbf{Read and recover}\\[2pt]$K_k=W_kM_k$\\[-1pt]{\footnotesize outcome $k=A,B$}};
\node[box,text width=2.42cm] (slot) at (8.35,0)
 {\textbf{External reward}\\[2pt]$r\sim\mathrm{Bern}(P_k)$\\[-1pt]{\footnotesize chosen arm $k$}};
\node[box,text width=2.64cm] (update) at (11.94,0)
 {\textbf{Reward update}\\[2pt]$\mathcal T_{k,r}$\\[-1pt]{\footnotesize retain or reset}};
\draw[flow] (state.east) -- (detect.west);
\draw[flow] (detect.east) -- (slot.west);
\draw[flow] (slot.east) -- (update.west);
\draw[flow] (update.north) -- ++(0,1.0) -| (state.north)
 node[pos=.31,above,font=\footnotesize]{next decision uses the updated preference};
\node[box,text width=3.02cm,fill=black!1] (avg) at (4.05,-3.02)
 {\textbf{Unselected state after $t$ trials}\\[2pt]$\bar\rho_t$\\[-1pt]{\footnotesize all outcome histories}};
\node[box,text width=3.6cm,fill=black!1] (diagnostic) at (9.63,-3.02)
 {\textbf{Separate fixed $C_5$ diagnostic}\\[2pt]$(\Pi_i,\Pi_{i+1})$\\[-1pt]{\footnotesize contextuality statistic $K_t$}};
\draw[probe] (state.south) -- node[left,pos=.57,font=\footnotesize,align=right]{many\\runs} (1.36,-3.02) -- (avg.west);
\draw[probe] (avg.east) -- node[above,font=\footnotesize]{fresh copies} (diagnostic.west);
\end{tikzpicture}
}
\caption{\textbf{Learning and diagnosis in one QTOW instrument.} The upper loop shows its repeated binary decision, external reward, and reward-conditioned update. Recovery $W_k$ preserves the logical preference plane; the update retains the state with weight $1-\gamma$ or resets it to $\tau_{k,r}$ with weight $\gamma$. The lower, dashed path represents an independent fixed $C_5$ test on fresh copies of the unselected learned ensemble $\bar\rho_t$. The diagnostic is an additional service, not a step in ordinary reward learning.}
\label{fig:qtow-operation-cycle}
\end{figure}

\subsection{Decision, observed reward, and the common state update}
\label{subsec:unified-instrument}
The instrument needs three physical modes but makes only \emph{two} decisions. We denote its basis by $|A\rangle,|B\rangle,|M\rangle$: modes $A$ and $B$ are associated with the competing arms, while $M$ is an intermediate mode that contributes to both detector outcomes. Mode $M$ is not a third slot machine. Within this three-dimensional space, two different superpositions $|\ell_+\rangle$ and $|\ell_-\rangle$ encode which arm is currently favored. These \emph{logical} states play the role of the two preference values $h=+1,-1$ in the classical model; they need not coincide with any single physical mode. Let $\mathcal H=\operatorname{span}\{|A\rangle,|B\rangle,|M\rangle\}$ be the real three-dimensional Hilbert space. We first describe the two-outcome readout, with Kraus operators
\begin{equation}
 M_A=\operatorname{diag}(1,0,1/\sqrt2),\qquad
 M_B=\operatorname{diag}(0,1,1/\sqrt2),\qquad
 K_k=W_kM_k\quad(k=A,B),
 \label{eq:unified-kraus}
\end{equation}
The diagonal operators $M_A,M_B$ describe the detector outcomes and satisfy $M_A^\dagger M_A+M_B^\dagger M_B=I$, so every decision produces one of the two arm labels. Their intermediate-mode entries $1/\sqrt2$ give that mode equal contributions to the two detector branches. A readout alone can displace the retained state from the plane used to encode preference. The fixed real rotations $W_A,W_B\in SO(3)$ undo that geometrical displacement: the recovered operators $K_k=W_kM_k$ return either outcome to the same logical plane $\mathcal C=\operatorname{span}\{|\ell_+\rangle,|\ell_-\rangle\}$. This is why recovery matters for learning: the next trial must act on the \emph{same kind of preference state}, independently of which arm was just selected. These are quantum-state recovery rotations, not rotations of the classical volume in Sec.~\ref{subsec:tow-principle}. The physical code states and rotations are specified constructively in Appendix~\ref{app:modular-recovery}. Write $P_\pm=|\ell_\pm\rangle\langle\ell_\pm|$ and $m=\operatorname{tr}[\rho(P_+-P_-)]$ for a diagonal logical state $\rho=(1+m)P_+/2+(1-m)P_-/2$. The restricted measurement satisfies
\begin{equation}
 K_A|\ell_\pm\rangle=\sqrt{\frac{1\pm \chi}{2}}|\ell_\pm\rangle,\qquad
 K_B|\ell_\pm\rangle=\sqrt{\frac{1\mp \chi}{2}}|\ell_\pm\rangle,
 \qquad \Pr(A\mid m)=\frac{1+\chi m}{2},
 \label{eq:unified-logical}
\end{equation}
where $0<\chi<1$ is the native decision contrast. The chosen arm $k\in\{A,B\}$ returns a Bernoulli reward $r\in\{0,1\}$ with $\Pr(r\mid k)=P_k^r(1-P_k)^{1-r}$. As in Sec.~\ref{subsec:classical-minimal-semantics}, $s_k$ is the arm-sign function ($s_A=+1$, $s_B=-1$), and $\chi$ is the dimensionless decision contrast introduced there. The indexed $P_k$ denotes the external reward probability $P_A$ or $P_B$; by contrast, $P_\pm$ are the two logical \emph{projectors} defined above and are not reward probabilities. The decision is not the end of the trial. After outcome $k$, the selected \emph{external} slot machine supplies the random reward $r$. A success on arm $A$ or a failure on arm $B$ supports the preference for $A$; a success on arm $B$ or a failure on arm $A$ supports the opposite preference. We encode these two possible lessons in the reset preparation $\tau_{k,r}=P_+$ if $s_k(2r-1)=+1$, and $P_-$ otherwise. The complete outcome-labelled, completely positive map is
\begin{equation}
 \mathcal T_{k,r}(\rho)=P_k^r(1-P_k)^{1-r}
 \left[(1-\gamma)K_k\rho K_k^\dagger+
 \gamma\operatorname{tr}(K_k\rho K_k^\dagger)\tau_{k,r}\right],\quad 0<\gamma<1.
 \label{eq:unified-trial}
\end{equation}
Equation~\eqref{eq:unified-trial} combines two familiar operations. With weight $1-\gamma$, the device keeps the recovered post-readout state, preserving what it had learned; with weight $\gamma$, it adopts the preference indicated by the new reward. The observed arm and reward determine \emph{which} state is reinforced, whereas $\gamma$ determines how strongly that single observation changes the retained preference. The branch factor $\operatorname{tr}(K_k\rho K_k^\dagger)$ accounts for the probability of observing arm $k$. Summing these outcome-labelled completely positive maps over all $k,r$ is trace preserving; conditioning on a particular observed $(k,r)$ gives the state used in the next real trial. Thus learning and readout form one repeated instrument, rather than independent mechanisms selected to obtain a desired final state. The reward update is a stochastic reset operation, \emph{not} a reward-conditioned unitary rotation. For native A/B decision--reward records, the instrument is exactly simulated by the classical two-state kernel of Eq.~\eqref{eq:cf-native-kernel}; the later contextuality witness tests a different, explicitly extended service.

\subsection{Stationary performance and the unselected ensemble}
\label{subsec:unified-adaptation}
For an individual run, the state depends on the sequence of selected arms and observed rewards. To ask what the \emph{learning process as a whole} produces, rather than selecting an unusually favorable sequence, we average over every possible branch with its actual probability. Starting with a mixture of the two logical preference states, this unconditioned average remains within that two-state family. The scalar $m_t$ measures its bias toward $|\ell_+\rangle$; $m_t=0$ means neither preference dominates, while positive $m_t$ means the ensemble increasingly favors arm $A$. With reward gap $\Delta=P_A-P_B$ and reward-probability sum $g=P_A+P_B$, the update gives
\begin{equation}
 m_{t+1}=[1-\gamma+\gamma \chi(g-1)]m_t+\gamma\Delta,
 \quad m_*=\frac{\Delta}{1-\chi(g-1)},\quad
 G_*=\frac{\chi\Delta^2}{2[1-\chi(g-1)]}>0\quad(\Delta\ne0),
 \label{eq:unified-learning}
\end{equation}
The first equation describes a feedback mechanism: if arm $A$ pays more often ($\Delta>0$), repeated successes move the average preference toward $A$; if the rewards are reversed, the bias changes sign. The final expression $G_*$ is the stationary reward advantage over choosing arms uniformly. It shows that the internal preference has a useful decision-making consequence, without asserting optimal regret. The algebra is provided in Appendix~\ref{app:unified-learning-calculation}. The physical state representing this average preference is
\begin{equation}
 \bar\rho_t=\frac{1+m_t}{2}P_++\frac{1-m_t}{2}P_-
 \label{eq:unified-ensemble}
\end{equation}
This \emph{unselected ensemble} includes successful and unsuccessful reward histories, not just the trajectories that happen to look quantum-like. It is produced by iterating the actual decision--reward map in Eq.~\eqref{eq:unified-trial}. The diagnostic below tests copies prepared after a specified number of trials; it does not alter the reward sequence used to reach that state.

\section{Contextuality Reached by the Same Decision Dynamics}
\label{sec:qtow-reachability}

\subsection{A fixed five-context diagnostic}
Reward learning and contextuality ask two different questions. The first asks whether experience improves the choice between $A$ and $B$. The second asks whether the \emph{same prepared state}, interrogated in several compatible ways, admits one context-independent meaning for the observed events. We therefore keep the diagnostic measurements fixed while the learned preparation evolves. The five tests form the $C_5$ cycle already introduced in Sec.~\ref{subsec:classical-gluing}; they are not five new reward choices. In particular, the native decision effects $E_k=M_k^\dagger M_k$ ($k=A,B$) are nonprojective, unsharp binary effects on the logical decision space for $0<\chi<1$. They must not be identified with the sharp, rank-one KCBS projectors $\Pi_i$ defined below. The five-context diagnostic uses separately controlled interrogations of repeated learned preparations, rather than treating the native A/B choice itself as a KCBS measurement. To realize this cycle in the present three-mode instrument, let $|n\rangle$ be the unit vector orthogonal to the two logical states. With $c^2=1/\sqrt5$, $s^2=1-c^2$, and $\varphi_i=4\pi i/5$, define before learning the five projectors
\begin{equation}
 |v_i\rangle=c|\ell_+\rangle+s\cos\varphi_i|\ell_-\rangle+s\sin\varphi_i|n\rangle,
 \quad \Pi_i=|v_i\rangle\langle v_i|,\quad i=0,\ldots,4.
 \label{eq:unified-projectors}
\end{equation}
The angle $\varphi_i$ specifies the fixed pentagonal probe orientation, $c^2=1/\sqrt5$ fixes its opening angle, and $|n\rangle$ is the third logical-orthogonal direction; these are diagnostic settings, not tunable reward probabilities. Adjacent rays are orthogonal, so $\{\Pi_i,\Pi_{i+1},I-\Pi_i-\Pi_{i+1}\}$ are valid three-outcome contexts. The same $\Pi_i$ is used in its two neighboring contexts. The five-ray sum is
\begin{equation}
 \Sigma=\sum_i\Pi_i=\sqrt5P_++\frac{5-\sqrt5}{2}(I-P_+).
 \label{eq:unified-KCBS-operator}
\end{equation}
Consequently, for the \emph{same, unselected} state of Eq.~\eqref{eq:unified-ensemble},
\begin{equation}
 \boxed{K_t=\operatorname{tr}(\bar\rho_t\Sigma)
 =\frac{5+\sqrt5+(3\sqrt5-5)m_t}{4},\qquad
 K_t>2\iff m_t>\frac1{\sqrt5}.}
 \label{eq:unified-KCBS}
\end{equation}
The formula makes the origin of the violation transparent. The diagnostic projectors, including their mutual exclusivities, do not change with trial number $t$; \emph{only the ensemble produced by reward learning changes}. At $m_0=0$, the identical test returns $K_0=(5+\sqrt5)/4<2$, so the chosen pentagon by itself does not make the initial state violate the bound. As rewarded decisions increase $m_t$, the fixed test crosses $K_t=2$ precisely when $m_t>1/\sqrt5$. This is a learning-generated \emph{state reaching a contextual region}, not a claim that the TOW update discovers the pentagonal measurement geometry. Access to the $C_5$ probe remains an additional declared function; native A/B choices are not themselves five KCBS contexts, and the construction does not establish minimal physical memory.

\subsection{Joint emergence of useful decisions and contextuality}
\begin{figure}[t]
\centering
\includegraphics[width=.87\linewidth]{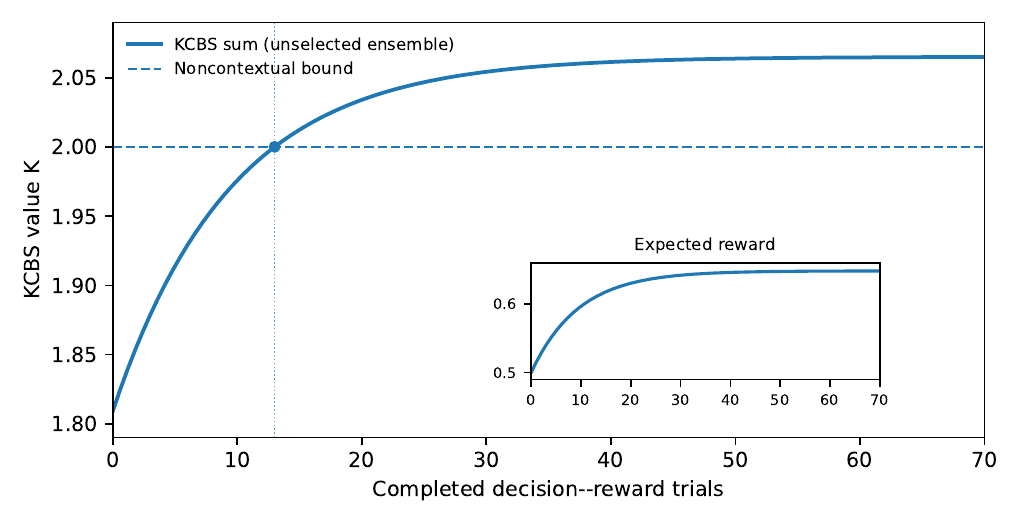}
\caption{\textbf{Same-instrument, unconditional learning trajectory.} The solid curve is the ideal sharp-probe KCBS sum of the ensemble after $t$ decision--reward trials; the dashed line is the noncontextual bound. The curve crosses the bound at $t=13$ and approaches $K_*=2.065248\ldots$. The inset gives the expected reward, which grows from the uniform-decision baseline $0.5$ to $0.648824\ldots$. Every trial branch is included.}
\label{fig:unified-trajectory}
\end{figure}

The following example shows both effects on the \emph{same time axis}. Set $P_A=0.8$ and $P_B=0.2$: arm $A$ is genuinely more rewarding, but the initially neutral agent has no preference. With the fixed physical recovery parameters $\sin\phi=9/16$, $\chi=\cos\phi=5\sqrt7/16$ and learning strength $\gamma=0.1$, take $m_0=0$. Each trial then shifts the average preference a little toward the more rewarding arm; iterating the update gives $m_t=0.6(1-0.9^t)$ and expected reward $R_t=0.5+(0.6\chi/2)m_t$. The reward advantage increases gradually, while the \emph{unchanged} C5 diagnostic detects a transition at trial $t=13$. In other words, the reward observations are responsible for moving the state across a threshold that the fixed tests can reveal. At stationarity,
\begin{equation}
 \boxed{R_*=0.648823511\ldots>0.5,\qquad
 K_*=2.065247584\ldots>2.}
 \label{eq:unified-joint-result}
\end{equation}
Figure~\ref{fig:unified-trajectory} and Table~\ref{tab:unified-trials} show the same unconditioned learning trajectory; detailed branch accounting is preserved in Appendix~\ref{app:unified-learning-calculation}. Both quantities refer to Eq.~\eqref{eq:unified-trial} without selecting particular rewards.

\begin{table}[t]
\centering\small
\caption{Decision and diagnostic results for one physical recovery--reset instrument, averaged over all reward histories.}
\label{tab:unified-trials}
\begin{tabular}{@{}rccc@{}}\toprule
Completed trials $t$ & Expected reward $R_t$ & Ideal KCBS $K_t$ & Finite-gate corrected gap \\\midrule
$0$ & $0.500000$ & $1.809017$ & $-0.195950$\\
$13$ & $0.610995$ & $2.000117$ & $-0.006320$\\
$14$ & $0.614777$ & $2.006630$ & $+0.000143$\\
$20$ & $0.630730$ & $2.034096$ & $+0.027397$\\
$30$ & $0.642515$ & $2.054386$ & $+0.047531$\\
Stationary & $\mathbf{0.648824}$ & $\mathbf{2.065248}$ & $\mathbf{+0.058309}$\\\bottomrule
\end{tabular}
\end{table}

\subsection{Finite, reused-operation contextuality test}
\label{subsec:unified-finite}
The sharp projectors above describe the clean theoretical test, but real reuse of the binary decision hardware gives only approximate implementations of the five contexts. This matters: a mismatch between two occurrences of a supposedly shared event should not be mistaken for contextuality. We therefore also consider a finite control program using \emph{the same physical instrument and its learned state}: the recovery rotations $W_A,W_B$ are independently called, and each context uses two binary interrogations. Seven explicit finite positive gate words generate five approximate three-outcome contexts; their four raw readout leaves are grouped as $AA$, $BB$, and $(AB\lor BA)$ (Appendix~\ref{app:modular-recovery}). Because occurrences of the same event in different contexts are not perfectly identical, let $u_i$ and $v_i$ be the two respective Yes probabilities of context $i$, and define
\begin{equation}
 K_{\rm av}(t)=\frac12\sum_{i=0}^4(u_i+v_i),\qquad
 \delta(t)=\sum_{i=0}^4|u_i-v_{i-1}|.
 \label{eq:unified-finite-mismatch}
\end{equation}
For the stated maximally coupled, consistently indexed noncontextuality criterion, the corrected bound is $K_{\rm av}\le 2+\delta/2$. Direct multiplication of the physical recovery matrices and seven finite words gives $K_{\rm av}(\infty)=2.067657790\ldots$, $\delta(\infty)=0.018696840\ldots$ and
\begin{equation}
 K_{\rm av}(\infty)-2-\delta(\infty)/2
 =0.058309370\ldots>0.
 \label{eq:unified-finite-gap}
\end{equation}
The correction subtracts the observed cross-context differences before declaring a violation. The finite-word excess becomes positive after $14$ learning trials, one trial later than for the ideal sharp diagnostic. The operational lesson is the same: an ordinary rewarded decision sequence prepares the state, and a separately calibrated diagnostic tests it. Calling the recovery gates independently, suspending reward reset during diagnostics, and preparing repeated copies are explicit additional capabilities of this \emph{one} instrument; they do not follow from the ordinary bandit interface alone. The finite-word test uses the Contextuality-by-Default (CbD) coupling convention; a separate preparation-noncontextual calibration witness using the same trained source is recorded in Supplementary Information (Sec. S3). An alternative weak-L\"uders/fixed-angle rotation example is retained for comparison in Supplementary Information (Sec. S8).

\section{From Reward Learning to an Operational Test of Contextuality}\label{sec:operational-comparison}
\subsection{What changes when a learned preference is interrogated in several contexts}
The preceding sections examine two ways of using a decision maker's internal preference. During ordinary TOW learning, that preference has one immediate role: it biases the next choice between arms $A$ and $B$. The selected arm produces a reward, and the reward changes the preference available for the following trial. Section~\ref{sec:classical-first} showed that a two-state classical process already performs this cycle. Section~\ref{sec:qtow-implementation} then implemented the same choice--reward task with a physical three-mode QTOW instrument. Its native records of choices and rewards can also be generated by the classical two-state process. In that limited task, the models are alternative realizations of the same observable learning behavior.

The question changes when the learned preference is used for more than choosing an arm. Suppose that the same retained state must also be interrogated through the five exclusive contexts of Fig.~\ref{fig:cf-c5-pentagon}. Each context asks about a pair of events, and each event reappears in a neighboring context. To describe these observations using one common event vocabulary, the meaning assigned to an event in one context must agree with its meaning in the other. This requirement connects probabilities that would otherwise be described independently. Unlike the reward probabilities $P_A$ and $P_B$, the five diagnostic probabilities are therefore subject to a cross-context consistency problem.

The classical analysis makes that problem quantitative. Equation~\eqref{eq:cf-semantic-cost} gives the smallest total mismatch needed to reconcile the five local event descriptions. When the mismatch is zero, one common noncontextual assignment is possible; a positive value means that at least some context-dependent identification is unavoidable. The important point is that the size of this mismatch is determined by the \emph{specified diagnostic task}. Holding the learning function and the information requirement fixed, we can ask how much common-event consistency can be retained. The calibrated information and GE2 conditions of Sec.~\ref{sec:classical-selection} identify a boundary at which the desired information can no longer be supplied with zero semantic repair. The resulting cost measures disagreement between event descriptions, not the number of bits needed to store a reward history.

The quantum construction provides the complementary dynamical result. Rather than imposing a state with a suitable KCBS value, the QTOW instrument repeatedly learns from the external slots. Its average state $\bar\rho_t$ develops a preference, and that evolving preparation is subsequently examined with the \emph{same fixed} five-context diagnostic. The crossing demonstrated in Sec.~\ref{sec:qtow-reachability} therefore belongs to a learning trajectory, although the ability to carry out the diagnostic is an additional, specified operation. Table~\ref{tab:operational-levels} shows how the successive tasks build on one another, from native decision making to shared-event information and its physical interrogation.

\begin{table}[H]
\centering\small
\caption{Successive operational tasks considered in the TOW framework. Each row adds a question about the same decision-making functionality rather than introducing a competing learning algorithm.}
\label{tab:operational-levels}
\begin{tabularx}{\linewidth}{@{}p{.23\linewidth}X p{.29\linewidth}@{}}\toprule
Model or service & Established result & Required capabilities\\\midrule
Two-state TOW learner & L1--L4 and stationary reward advantage & Coupled preference, choice and reward update\\
Five-context information service & Exact minimal semantic repair and fixed-information boundary; conditional information-optimal selection & Shared events, neutral calibration, information requirement and GE2\\
Two-copy calibrated diagnostic & Quantum-compatible maximal $C_5$ vector under GE2 and efficiency & Independent copies and adjustable response calibration\\
Recovery--reset QTOW & Stationary gain and unconditional ideal KCBS violation; finite corrected CbD witness & One native learning instrument; independent diagnostic reuse of recovery gates\\
\bottomrule
\end{tabularx}
\end{table}

\subsection{Following the learned state into a contextuality measurement}
The distinction between learning and probing also suggests how to examine the prediction experimentally. First, identical devices are allowed to undergo a specified number $t$ of genuine bandit trials, with the actual choices and rewards determining their updates. Repeating this procedure prepares independent copies of the \emph{unselected} ensemble $\bar\rho_t$: no particular reward sequence is retained or discarded to improve the diagnostic result. A context is then chosen at random, and its three mutually exclusive outcomes are recorded. By repeating the preparation and measurement for each of the five adjacent pairs $(\Pi_i,\Pi_{i+1})$, one estimates the KCBS expression of Eq.~\eqref{eq:unified-KCBS}. In this way, the horizontal axis of the learning trajectory has an operational meaning: it is the number of reward experiences accumulated before asking the same five diagnostic questions.

The ideal calculation uses projectors that are genuinely shared by their two neighboring contexts. In an apparatus, however, the same event label does not guarantee identical responses when the event is implemented within different measurement sequences. This is why the finite-control construction of Sec.~\ref{sec:qtow-reachability} is useful. It replaces freely specified ideal rotations by finite sequences of the instrument's recovery operations, and compares the two measured occurrences of each event. Any differences in their marginals enter the corrected bound of Eq.~\eqref{eq:unified-finite-mismatch}; a positive corrected margin then has a clear meaning even when the implementations are not exactly identical. Calibration, compatibility and preparation stability remain part of the measurement protocol, as emphasized in discussions of finite-precision contextuality tests \cite{Meyer1999,BarrettKent2004} and of context-dependent marginals \cite{dzhafarov2015contextuality,DzhafarovKon2018}.

The same distinction between preparation and interrogation suggests a possible behavioral test, although its measurement assumptions would have to be validated independently. For example, a prospective decision experiment could predefine five binary judgment events $e_0,\ldots,e_4$ and present the five neighboring pairs $(e_i,e_{i+1})$ in randomly assigned contexts. In each context, the task would have three explicitly distinguished possible reports---event $e_i$, event $e_{i+1}$, or neither---so that simultaneous positive reports for adjacent events are excluded by the protocol. To interpret the resulting frequencies as a KCBS experiment, however, the study would still have to establish that the same event has the same operational meaning in its two contexts, that a paired interrogation is sufficiently compatible or nondisturbing, and that preparations can be repeated comparably. The three-response format alone would not establish any of these nontrivial conditions. Marginal equality could be checked directly; if it fails, a suitably specified context-indexed analysis would be required. Such a design is an experimental proposal, not a validated set of psychological KCBS measurements, and no human KCBS violation is claimed here.

The difficulty of satisfying these conditions in behavioral experiments also makes the choice of contextuality test important. Bell-type tests use correlations across alternative settings, typically on separated or otherwise distinctly indexed subsystems; the ideal KCBS test instead uses five pairwise compatible, exclusive events on one system. Contextuality-by-Default (CbD) assigns separate random variables to each occurrence of an event and assesses whether couplings consistent with the observed within-context distributions can identify those occurrences, allowing for unequal marginals (``inconsistent connectedness''). For cyclic systems, generalized Bell-type inequalities and measures can be formulated within this framework \cite{dzhafarov2015contextuality,DzhafarovKujalaCervantes2020}. Our finite-control criterion in Sec.~\ref{subsec:unified-finite} similarly prices cross-context marginal differences, but is applied here only to the specified five-cycle and its stated coupling convention; it should not be identified with every form of Bell test or every possible CbD analysis. Thus neither a question-order effect nor a violation of another inequality by itself replaces the compatibility and exclusivity tests needed for our KCBS construction.

A complementary test probes a different part of the same operational service. Supplementary Information, Sec.~S3, constructs preparation mixtures that are operationally equivalent and derives a calibrated preparation-noncontextuality inequality. That test asks whether differently prepared yet operationally equivalent mixtures can share one underlying preparation description. It uses a different hypothesis from the finite-control CbD analysis, but it serves the same overall purpose: translating the abstract requirement of common meaning into observable preparation and measurement relations. A photonic qutrit implementation route is outlined in Supplementary Information, Sec.~S7.

\subsection{Relating the representations to physical operations}
The examples above also show why a physical realization matters when discussing operational minimality. In the original TOW picture, conservation couples two volumes and creates a shared preference variable; fluctuations and external rewards drive its evolution. In the QTOW instrument, state normalization and the recovery maps instead make repeated binary readout compatible with retention of a logical preference. These are different physical mechanisms, but both let information acquired from one choice affect the next choice through a common state. The semantic cost studied in Secs.~\ref{sec:classical-first}--\ref{sec:classical-selection} characterizes a further requirement: whether that state can support several mutually connected event descriptions without additional context-dependent distinctions.

These parallels should not obscure the additional controls required for the diagnostic. To perform the idealized five-context comparison, the instrument must provide independently selectable diagnostic settings; to implement the finite version, its recovery gates must be callable outside the ordinary reward-update cycle, and the reward reset must be suspended during probing. These controls are specified explicitly so that the two tasks can be compared on a common experimental platform. Other QTOW realizations can use different feedback mechanisms: Supplementary Information, Secs.~S8--S9, analyzes a fixed-angle protocol and its finite-time behavior, while Sec.~S6 considers asymmetric rotations. Their physical differences help delimit which conclusions follow from the shared decision function and which depend on a particular implementation.

The need to account for these operations is also central to a fair comparison with classical models. A ray-state sampler, for example, can reproduce Born-rule probabilities and conditional state updates without necessarily satisfying preparation noncontextuality \cite{HarriganSpekkens2010}. Reproducing the statistics and preserving the specified shared-event identities are therefore different requirements. The tests above state their operational equivalences and noncontextuality hypotheses explicitly.\par
A related concern with preserving useful information under representational constraints appears in recent AI research, though the optimized quantities are different. In-context learning viewed through rate--distortion optimization selects and compresses demonstrations within a token budget while limiting predictive distortion \cite{ZhangLiXiao2026RD}; here the scarce resource is a prompt representation and the performance criterion is predictive quality. Replay-free continual multimodal learning has also been studied using compact prompt-invariant canonical-correlation certificates to preserve cross-modal alignment without storing past training examples \cite{ZhangEtAl2026PiCCA}; that setting concerns retention of learned alignment under distribution and prompt changes. In contrast, our $D_{\mathrm{sem}}^{\min}$ is the minimum probability of disagreement required to relate repeated occurrences of five shared events under the stipulated $C_5$ service. It is neither a token budget nor a bound on stored model parameters or physical memory. The comparison thus illuminates a common question about reusing representations across tasks, without equating information compression with semantic repair.

In summary, the native A/B learning history, the shared five-context information service, and the learned-state QTOW diagnostic impose successively different operational demands. Their comparison concerns the consistency of reused event meanings, not an unmatched claim of memory or computational advantage. The Discussion draws together these complementary results.

\section{Discussion and Conclusion}\label{sec:conclusion}
Our starting point is a \emph{fixed} operational task: useful reward-adaptive choices together with a specified capacity to retrieve information from shared internal preferences. The native TOW functionality admits a fully classical two-state model. When that same device is required to support five calibrated, exclusive shared-event contexts, however, context-local classical probabilities need not admit one joint, context-independent event assignment. We establish the exact least repair $D_{\rm sem}^{\min}=2[K-2]_+$ for the ideal $C_5$ service. Within the stated calibrated, GE2-admissible diagnostic family, fixing an information requirement and minimizing this semantic repair gives the same efficiency boundary as maximizing accessible diagnostic information under a repair budget. Above its computed threshold, preserving the required information entails a strictly positive failure of common-event identity. This is an operationally defined semantic minimum, not a claim of minimum physical storage or a universal deduction of contextuality from reward learning alone.

A separate, constructive result realizes the learned-state statistics in one quantum decision instrument. The same recovered binary readout and externally rewarded reset update yield both stationary gain over uniform decisions and an unconditional learned state crossing the ideal KCBS bound after thirteen trials, approaching $K_*=2.065248\ldots$. The recovery controls of that instrument can also be used in a separate finite diagnostic, where the contextuality gap remains positive after correcting for differences between event occurrences. Thus the contextuality witness does not rely on an externally selected state or a favorable postselected reward history.

\paragraph{Connection with quantum cognition and natural/artificial intelligence.}
Quantum-cognitive models use noncommuting judgments to organize order effects and other context-sensitive behavioral statistics \cite{busemeyer2012quantum,wang2014order}. Our result addresses a different level of explanation: it is a construction and a representational boundary for a specified decision device, not a fit to human-subject data or evidence for quantum physics in the brain. A psychological KCBS test would additionally have to establish compatible shared events, exclusivity and consistent connectedness, or use an appropriate corrected contextuality criterion when measured marginals differ \cite{dzhafarov2015contextuality}. In particular, an ordinary question-order effect cannot be identified with a KCBS violation merely by changing the order of two questions.

For the distinction between natural and artificial intelligence, the relevant contribution is the location of contextual information. A physically constrained decision mechanism can reuse a common internal state across learning and compatible interrogations; an artificial probabilistic program can also reproduce context-labelled observations by representing preparation, context and history explicitly. Nothing in the KCBS witness proves that one architecture is universally faster, smaller, or more intelligent. Rather, the calculation identifies the price of retaining a \emph{single noncontextual meaning} for the prescribed five shared events, while the constructed QTOW instrument illustrates how contextual statistics can coexist with a concrete reward-learning process. This sharpens the architectural question raised in quantum-like approaches to natural and artificial intelligence \cite{Khrennikov2026}.

The results make a precise distinction between a shared operational event description and unrestricted classical emulation. Context-aware stochastic simulations remain possible; a noncontextual assignment respecting the specified identities and exclusivities does not, once the boundary is crossed. The classical-first repair theorem, the conditional information-efficiency selection, and the same-instrument QTOW realization together show how maintaining fixed decision and information functions with economical common semantics can expose contextual probability. The Supplementary Information examines complementary weak-readout and fixed-angle protocols with distinct update dynamics, allowing their results to be compared without conflating them with the recovery--reset instrument. For quantum cognition, this connects a well-defined contextuality obstruction to a learning-generated preparation, while leaving open whether comparable operational contexts occur in human judgments. For artificial intelligence, it frames the comparison in terms of operational state reuse and explicit contextual bookkeeping rather than an unqualified claim that human decision making follows quantum mechanics.

\section*{Acknowledgements}

The author thanks Professor Taiki Takahashi (Hokkaido University) for early discussions related to this work.

\section*{Declaration of generative AI and AI-assisted technologies in manuscript preparation}

During the preparation of this work, the author used ChatGPT for drafting and language editing, manuscript organization, and assistance with numerical verification scripts. Responsibility for the final scientific arguments, the verification of results, and the manuscript rests with the author.

\section*{Declarations}

\subsection*{Ethical approval}
Not applicable. This study is theoretical and computational; no human or animal participants were involved.

\subsection*{Consent to participate}
Not applicable.

\subsection*{Consent to publish}
Not applicable. This manuscript contains no identifiable individual participant information.

\subsection*{Funding}
This work was supported by SOBIN Institute LLC under Research Grant SP004.

\subsection*{Competing Interests}
The author is a Guest Editor of the Collection ``Difference between Artificial Intelligence and Natural Intelligence'' in \emph{Discover Artificial Intelligence}. The author will not participate in the editorial handling, peer-review process, or publication decision for this manuscript. The author declares no other competing interests.

\subsection*{Data Availability Statement}
No empirical human-participant or device measurements were collected. The numerical trajectories and probability tables were generated from the equations and parameters specified in the article. The accompanying reproducibility package contains the numerical outputs, input parameters, and verification scripts; these materials are also available from the corresponding author upon reasonable request.

\subsection*{Materials availability}
Not applicable.

\subsection*{Code availability}
The accompanying reproducibility package contains the simulation and independent checking scripts used for the reported theoretical and numerical results.

\subsection*{Author contribution}
The author solely conceived the study, developed the theoretical framework, performed the analysis, and wrote the manuscript.

\begin{appendices}
\section{Supporting classical-first derivations}
\label{app:classical-first}

\subsection{Constructive classical repair of the exclusive pentagon}

Let $p_i\ge0$ satisfy $p_i+p_{i+1}\le1$ and let
$K=\sum_i p_i$. If $K\le2$, the vector $p$ belongs to
$\operatorname{STAB}(C_5)$ and is the mean of an ordinary
probability distribution on the eleven independent sets of the
pentagon. Thus all shared events admit a single global joint model.

If $2<K\le5/2$, put $\alpha=2(K-2)$.
The disjoint edge cover of the four vertices other than $i$
implies $p_i\ge K-2=\alpha/2$.
For $\alpha<1$ define
$r_i=(p_i-\alpha/2)/(1-\alpha)$. Then
$r_i\ge0$, $r_i+r_{i+1}\le1$, and $\sum_i r_i=2$.
Such a point belongs to the convex hull of the five size-two
stable-set indicators; e.g. one can take
$u_i=1-r_{i+3}-r_{i+4}$ and write
$r_i=u_i+u_{i-2}$, with $u_i\ge0$ and $\sum_i u_i=1$.

Splitting only event $e_0$ into its two context occurrences
turns the pentagon into a path with six vertices. Assign total
weight $1-\alpha$ to the common stable-set assignments,
in the proportions $u_i$, and total weight $\alpha$ equally
to the two alternating assignments of this six-vertex path.
The alternating assignments disagree only at the two copies
of $e_0$. Their mixture has the required five exclusive
pair distributions and total mismatch $\alpha=2(K-2)$.
For $\alpha=1$ use only the two alternating assignments.
This establishes the attainability claimed in
Eq.~\eqref{eq:cf-semantic-cost}.

\section{Functional selection of a contextual diagnostic: proof and classical controls}
\label{app:functional-selection}

This appendix proves the conditional selection theorem of Sec.~\ref{subsec:classical-selection-functional} within the displayed response family. It uses only positive context-local classical probabilities, Shannon mutual information and independent-copy event exclusivity. These assumptions are \emph{additional tasks}, not consequences of L1--L4.

\subsection*{Response validity and two-copy exclusivity}
Put $A=1/3+\kappa/2$, $\epsilon=1/2000$, $d=(-2,-1,0,1,2)$. The Yes response of event $i$ is $r_i(h)=1/3+\epsilon d_i+\kappa h/2$, and the three outputs of context $c_i$ have probabilities $(r_i(h),r_{i+1}(h),1-r_i(h)-r_{i+1}(h))$. For $0\le\kappa\le\kappa_{\mathrm{GE}}$, all probabilities are nonnegative. The coarse-grained Yes and No instruments act by diagonal stochastic multiplication on $h$ and are identical in the two appearances of event $i$.

For the OR-product exclusivity graph of two pentagons, the ten maximal cliques of size five project bijectively onto both pentagon copies, and their offsets obey $\sum_{(i,j)\in Q}d_i d_j\in\{-5,0,5\}$. The largest clique sum over independently preparable states is therefore $5A^2+5\epsilon^2$, giving precisely
\[
 5(1/3+\kappa/2)^2+5\epsilon^2\le1
 \quad\Longleftrightarrow\quad
 \kappa\le2\bigl(\sqrt{1/5-\epsilon^2}-1/3\bigr).
\]
All maximal cliques with fewer than five vertices obey the inequality strictly because $4(A+2\epsilon)^2<1$ over this domain; the checks extend to the allowed product preparations since each Yes response is largest at $h=+1$. A reproducible enumeration of the 25 vertices of the OR-product exclusivity graph and all its 535 maximal cliques (including ten cliques of size five) is implemented in the accompanying \texttt{verification/verify\_classical\_functional\_selection.py}. The script checks every resulting clique inequality for the allowed response family; its exact integer graph-adjacency and offset sums underpin the finite classification. This is a constraint on an explicitly \emph{admitted preparation family}, not a proof that an arbitrary classical diagnostic automatically has GE2.

\subsection*{Information-optimal contrast}
Let $H\in\{\pm1\}$ be uniform and let context $C$ be independently uniform on the five choices, and $O\in\{10,01,00\}$ be the corresponding observed outcome. For $0\le\kappa_1<\kappa_2\le\kappa_{\rm GE}$, the output channel at $\kappa_1$ is reproduced by observing the channel at $\kappa_2$, retaining its result with probability $\kappa_1/\kappa_2$, and otherwise drawing a fresh result from the $\kappa=0$ distribution for the same context. This is a stochastic garbling independent of $H$. The data processing inequality gives $I(H;O\mid C)_{\kappa_1}\le I(H;O\mid C)_{\kappa_2}$, strictly when $\kappa_1<\kappa_2$ under the positive nontrivial contrasts. Hence maximizing diagnostic informativeness chooses the upper boundary $\kappa_{\rm GE}$.

Since $\sum_i d_i=0$ and $m=\mathbb E[H]$, summing the five marginals gives $K(m)=5/3+(5\kappa_{\rm GE}/2)m$. Using $m_t=0.6(1-0.9^t)$ from the native model when $P_A+P_B=1$ gives $K_*=2.008306614640\ldots$ and first violation at $t=36$. This is an exact classical stochastic process at every stage; the obstruction concerns a \emph{single context-independent exclusive event-value assignment}, not Kolmogorov sampling of experiments with context as an input.

\subsection*{Why the extra design requirements matter}
For any context-local instrument, let an independent Yes-to-neither reporting loss have efficiency $0\le t\le1$: it sends each Yes output to ``neither'' with probability $1-t$. It leaves all decision--reward histories and shared-event identification unchanged but replaces every $p_i$ with $tp_i$, hence $K\mapsto tK$; under two-copy exclusivity the product Yes weights only decrease. Thus no set of axioms indifferent to this detection efficiency can \emph{force} $K>2$. Moreover, the different family $r_i(-)=a+\epsilon d_i$, $r_i(+)=a+1/9+\epsilon d_i$ admits a diagnostically informative noncontextual calibration at $a=1/4$ with $K(m_*)=61/36<2$ for the same bandit learner; the contextual calibration at $a=1/3$ is therefore not forced by native learning or shared event labels alone. Maximizing information while allowing the offset $a$ to vary can itself prefer a noncontextual boundary, so the neutral calibration in the principal theorem is substantive. These are constructive controls, not deficiencies of the conditional selection result.

\section{Full composition and calibration derivations}\label{app:calibration-details}
We derive the GE2 constraints and the efficient calibration family used in Sec.~\ref{sec:classical-selection}, with the required composition and preparation assumptions stated explicitly.
\subsection*{Quantum-compatible C5 selection from composition and calibration}

\subsubsection*{Independent composition and global exclusivity}

The local exclusivity constraints by themselves allow
$K\le5/2$, whereas a common, predetermined noncontextual assignment
requires $K\le2$. We now ask which \emph{additional operational
requirements}, formulated without Hilbert-space probabilities,
select the familiar quantum KCBS upper value.
This subsection considers a composition service \emph{independent of}
the minimal TOW learning specification.

Take two independently prepared copies of the same five-event system.
For the joint Yes events $F_{ij}=e_i^{(1)}\wedge e_j^{(2)}$, assume
$\Pr(F_{ij})=p_i p_j$. Impose two-copy global exclusivity (GE2):
the sum of probabilities of any mutually exclusive set of these
product Yes events is at most one. Exclusivity of two product events
is guaranteed whenever one factor contains an exclusive pair.

\noindent\textbf{Proposition 3 (GE2 upper bound).}
For any local-exclusive $C_5$ preparation satisfying this independent
two-copy GE2 condition,
\begin{equation}
 \boxed{\displaystyle K\le\sqrt5.}
 \label{eq:appcf-ge2-k}
\end{equation}
\emph{Proof.} For each $s=0,\ldots,4$, the five product events
$\{F_{i,\,2i+s}:i=0,\ldots,4\}$ (indices mod five) form a clique of
mutually exclusive events. Therefore
$\sum_{i=0}^4 p_i p_{2i+s}\le1$.
The five cliques partition the 25 ordered pairs $(i,j)$; summing
their inequalities yields $(\sum_i p_i)^2\le5$.
\hfill$\square$

This classical-probability derivation of the quantum KCBS upper bound
from global exclusivity is \emph{not} claimed as new; see
Cabello's global-exclusivity argument and the graph-theoretic
framework of Cabello, Severini, and Winter
\cite{Cabello2013GE,CabelloSeveriniWinter2014}.
GE2 is a genuine independent assumption about combined experiments,
not a consequence of the five separate local tables. A preparation
$p_i=1/\sqrt5$ saturates Eq.~\eqref{eq:appcf-ge2-k} in the symmetric
class, but GE2 alone neither requires its preparation nor establishes
the Born rule.

\subsubsection*{Efficient calibration from a two-copy probability budget}
\label{app:efficient-calibration}

Consider $r_i(\vartheta)=\tau(\vartheta)+\vartheta v_i$ with $v=(2,-1/2,-1/2,-1/2,-1/2)$ and $0\le\vartheta\le h=10^{-2}$. We treat the possibility of continuously adjustable calibration as an additional operational service. The clique of product events $F_{i,2i}$ gives
\begin{equation}
\sum_i r_i(\vartheta)r_{2i}(\vartheta)=5\tau(\vartheta)^2+5\vartheta^2\le1,
\label{eq:app-efficient-envelope}
\end{equation}
where $\sum_i v_i=0$ and $\sum_i v_i v_{2i}=5$. At fixed $\vartheta$, maximizing $K(r)=5\tau$ over the positive branch of this self-composition constraint yields the efficient curve $\tau_{\rm eff}(\vartheta)=\sqrt{1/5-\vartheta^2}$, Eq.~\eqref{eq:cf-efficient-cal-family}. This is a constrained efficiency choice for a diagnostic response, not a claim that native TOW learning must optimize $K$.

\noindent\textbf{Proposition (GE2 compatibility of the efficient family).}
For $\vartheta,\varphi\in[0,h]$, all local probabilities $r^{\rm eff}$ are nonnegative and adjacent pair sums are at most one. Five representative size-five product cliques, indexed by $s\in\{0,1,2,3,4\}$, satisfy
\begin{equation}
\sum_i r^{\rm eff}_i(\vartheta)r^{\rm eff}_{2i+s}(\varphi)
=5\tau_{\vartheta}\tau_{\varphi}+S_s\vartheta\varphi\le1,
\quad S_0=5,\quad S_{s\ne0}=-5/4.
\label{eq:app-efficient-ge2-cliques}
\end{equation}
Indeed $S_s\le5$, and Cauchy--Schwarz gives
$5(\tau_{\vartheta}\tau_{\varphi}+\vartheta\varphi)\le
5\sqrt{\tau_{\vartheta}^2+\vartheta^2}\sqrt{\tau_{\varphi}^2+\varphi^2}=1$.
The transposed size-five cliques obey the analogous bound. Smaller product cliques satisfy GE2 because, in this restricted range, each response is bounded by $1/\sqrt5+2h<1/2$; the remaining clique inequalities can be verified by the finite exclusivity-graph enumeration supplied with the calibration calculation. The same construction is compatible with the stable-set preparations. These statements establish a consistent example of the assumed calibration service.

\noindent\textbf{Proposition (rigidity on the $K=\sqrt5$ face).}
Let a preparation $p$ satisfy $\sum_i p_i=\sqrt5$ and be mixed-GE2 compatible with every $r^{\rm eff}(\vartheta)$ for sufficiently small $\vartheta>0$. With $t=1/\sqrt5$ and $i_s$ uniquely defined by $2i_s+s\equiv0\pmod5$, the product-clique constraint yields
\[
\sum_i p_i r^{\rm eff}_{2i+s}(\vartheta)
=\sqrt{1-5\vartheta^2}+\frac52\vartheta(p_{i_s}-t)\le1.
\]
Hence
\begin{equation}
 p_{i_s}-t\le
 \frac{2\vartheta}{1+\sqrt{1-5\vartheta^2}},
 \label{eq:app-efficient-rigidity-bound}
\end{equation}
which vanishes as $\vartheta\downarrow0$. Thus $p_i\le t$ for every $i$ and $\sum_i p_i=5t$ forces $p_i=t$ for all $i$. In this precise sense, the efficient calibration identifies the uniform quantum-compatible maximal point by classical GE2 inequalities plus independently granted calibration access.

If the relevant mixed GE2 inequality holds only within an additive tolerance $\delta$, the right-hand side of Eq.~\eqref{eq:app-efficient-rigidity-bound} acquires an additional term $2\delta/(5\vartheta)$. At finite resolution, it is therefore an approximate selection result. The two-copy constraint is not a derivation of quantum measurement probabilities for arbitrary observables.

\subsubsection*{Scope and the role of QTOW}

GE2 and the efficient calibration service determine a bound and, under their respective assumptions, a distinguished
\emph{quantum-compatible KCBS maximal probability vector}.
They do \emph{not} reconstruct the quantum probability set,
Hilbert-space geometry, the Born rule, or uniquely quantum
dynamics. Other nonquantum empirical models can satisfy
GE2 and less demanding calibration requirements;
nontrivial calibration access is therefore essential.

There are three distinct claims in the present development.
First, native classical reward learning is sufficient for the
ordinary decision histories. Second, classical context-local
probabilities subject to shared-event semantics can exhibit a
quantifiable failure of noncontextual gluing, conditionally on
the diagnostic service. Third, further composition and
calibration requirements can select a quantum-compatible
numerical extremal point. These results are complementary:
none is silently substituted for a derivation of the others.

Only \emph{after} establishing these classical-probability
boundaries do we adopt the qutrit state space and generalized
quantum decision instrument of Sec.~\ref{sec:qtow-implementation}
as an explicit, compact \emph{realization}. The qutrit
construction uses Hilbert-space structure and Born/Kraus
probabilities as modeling assumptions; its additional result
must therefore concern operational implementation and
reward-driven \emph{reachability} of contextual states,
not a claimed derivation of Born probabilities from TOW alone.

\section{Learning recurrence and finite-contextuality bound}\label{app:unified-learning-calculation}
The outcomes of the physical recovery--reset instrument on the logical states $P_{+}$ and $P_{-}$ have the same event-labelled Markov kernel as Eq.~\eqref{eq:cf-native-kernel}. Conditionally on preference $h=\pm1$, choice $a$ and reward $r$, the expectation of the next preference is
\begin{equation}
 \mathbb E[h'\mid h,k,r]=(1-\gamma)h+\gamma s_k(2r-1).
\end{equation}
Averaging the Bernoulli reward gives $\mathbb E[2r-1\mid k]=2P_k-1$. Using $\Pr(k\mid h)=(1+s_k \chi h)/2$ then yields
\[
 \mathbb E[s_k(2r-1)\mid h]=P_A-P_B+\chi(P_A+P_B-1)h.
\]
Finally, averaging $h$ derives Eq.~\eqref{eq:unified-learning}. The expected reward is $R_t=(P_A+P_B)/2+\chi(P_A-P_B)m_t/2$, giving the stationary gain shown there. The native recursion therefore includes \emph{all} choice--reward branches, rather than postselecting a favorable reward sequence.

For the fixed KCBS rays of Eq.~\eqref{eq:unified-projectors}, adjacent orthogonality follows from $c^2+s^2\cos(4\pi/5)=0$. The trigonometric pentagon sums give $\sum_i\Pi_i=\sqrt5 P_++(5-\sqrt5)(I-P_+)/2$, hence Eq.~\eqref{eq:unified-KCBS} on the same unselected ensemble. At the stated parameters $m_t=0.6(1-0.9^t)$; substitution yields the first ideal $K_t>2$ at $t=13$.

For a finite diagnostic, write its two incident occurrences of event $i$ as binary variables $L_i,R_i$. Each context has mutually exclusive Yes outcomes. For any single global coupling of their ten outcomes, this implies the pathwise five-cycle inequality
\[
 \frac12\sum_{i=0}^4(L_i+R_i)\le 2+\frac12\sum_{i=0}^4|L_i-R_i|.
\]
Under a noncontextual maximally connected coupling, each occurrence pair has the minimum mismatch probability $|\Pr(L_i=1)-\Pr(R_i=1)|$; taking expectations gives the corrected CbD-type bound $K_{\rm av}\le2+\delta/2$ used in Sec.~\ref{subsec:unified-finite}. This is a bound conditional on the specified coupling notion, not the distinct preparation-noncontextuality bound. The finite recovery words and direct matrix evaluation are recorded in Appendix~\ref{app:modular-recovery}.

\section{Recovery instrument, finite control words, and noncontextuality calibration}\label{app:modular-recovery}
The same recovery-and-reset qutrit instrument is used in the native learning theorem of Sec.~\ref{sec:qtow-implementation} and the contextuality diagnostics of Sec.~\ref{sec:qtow-reachability}. Only the access commands change between the learning and diagnostic modes. The former uses composite recovered readout $K_k=W_kM_k$ followed by the reward-conditioned reset of Eq.~\eqref{eq:unified-trial}; the latter allows independent calls to the already existing physical recovery rotations and suppresses the ordinary bandit reward/reset during diagnostic interrogations.

\subsection*{Physical realization of the logical decision state}
Let $A,B,M$ be unit coordinate vectors in $\mathbb R^3$ and put $p=\sqrt{(1+\chi)/2}$, $q=\sqrt{(1-\chi)/2}$. A real orthonormal logical code and auxiliary unit vector are
\begin{equation}
 \ell_+=p^2A+q^2B+\sqrt2pqM,\qquad
 \ell_-=q^2A+p^2B-\sqrt2pqM,\qquad
 n=\ell_+\times\ell_-.
 \label{eq:recovery-code}
\end{equation}
The ranges of the physical readout $M_A,M_B$ have orthonormal frames
\begin{align}
 j_{A+}&=pA+qM,& j_{A-}&=qA-pM,\\
 j_{B+}&=qB+pM,& j_{B-}&=pB-qM.
\end{align}
With $O=(\ell_+,\ell_-,n)$, $O_A=(j_{A+},j_{A-},B)$ and $O_B=(j_{B+},j_{B-},-A)$, choose $W_A=OO_A^T$ and $W_B=OO_B^T$. Both matrices lie in $SO(3)$. A direct multiplication verifies
\begin{equation}
 K_A\ell_\pm=\sqrt{(1\pm \chi)/2}\,\ell_\pm,\qquad
 K_B\ell_\pm=\sqrt{(1\mp \chi)/2}\,\ell_\pm,
 \label{eq:recovery-identity}
\end{equation}
which proves the logical decision law Eq.~\eqref{eq:unified-logical} without using a new measurement effect. The full operation for observing action $k$ and reward $r$ is Eq.~\eqref{eq:unified-trial}, including the probability of the external reward. Its two quantum branches are completely positive and the sum over all $k,r$ is trace preserving. On diagonal logical states, the recovered measurement has the same native event-labelled probability kernel as the classical Markov learner of Eq.~\eqref{eq:cf-native-kernel}. This equivalence does not extend automatically to the separate five-context diagnostic service.

\subsection*{Finite contexts from positive recovery words}
For the parameter $\chi=5\sqrt7/16$, seven positive words in the same rotations $W_A,W_B$ give a finite diagnostic. Products are applied in the written matrix order, from left to right; no inverse gate is invoked. Two words are intermediate corrections $D_A,D_B$ and five are pre-context rotations $U_i$. The letter strings specifying the products are:
\begin{quote}\ttfamily\small
D\_A = AAABBAAABBAAAAAAABAA\\
D\_B = BABAAAAAABBAABBBAAABAAAA\\
U\_0 = AABBBBABBABBABBBBAA\\
U\_1 = AABABABABABBABBABBAB\\
U\_2 = ABBBABABAABBABBBBAAABA\\
U\_3 = AABAAAAABABAAAABBAAAAAB\\
U\_4 = ABAAABBABBABBBBBAABABB
\end{quote}
The four raw Kraus products for the two-interrogation protocol in context $i$ are $H_{k_1 k_2}U_i$, where $H_{k_1 k_2}=K_{k_2}D_{k_1}K_{k_1}$ and $k_1,k_2\in\{A,B\}$ label the first and second binary outcomes, respectively. Grouping their effects as $AA$, $BB$ and $AB\lor BA$ gives a positive normalized three-outcome POVM. The first and second Yes effects for the five contexts need not coincide exactly between neighboring contexts. On the unselected learned state $\bar\rho_t$ of Eq.~\eqref{eq:unified-ensemble}, compute each occurrence probability and the correction $\delta(t)$ defined in Eq.~\eqref{eq:unified-finite-mismatch}. The resulting stationary values are
\begin{equation}
 K_{\mathrm{av}}=2.067657789899\ldots,\quad
 \delta=0.018696840205\ldots,\quad
 K_{\mathrm{av}}-2-\delta/2=0.058309369797\ldots.
 \label{eq:recovery-finite-values}
\end{equation}
They result from direct multiplication of the explicit physical $3\times3$ matrices and finite words; the complete finite-time calculation is provided in the accompanying reproducibility script. The corrected excess is positive at $t=14$. These finite-word tests use ideal real-valued control specifications and do not certify finite-bit experimental tolerances.

\subsection*{Operational boundary}
The recovery gates $W_k$ must be callable independently, the bandit reward reset must be suspendable for diagnostic interrogations, and prepared copies of $\bar\rho_t$ must be stable. These are additional services of the \emph{same} physical instrument, not consequences of being able to choose between arms $A$ and $B$ during native learning. The ability to represent all ordinary choice--reward histories classically is consistent with the failure of a more restrictive noncontextual description of those extended diagnostic preparations and events. The separate preparation-noncontextuality construction and the alternative weak-L\"uders/fixed-angle models are retained in Supplementary Information.

\end{appendices}

\clearpage
\phantomsection
\pdfbookmark[0]{Supplementary Information}{supplementary-information}
\begingroup
\setcounter{section}{0}
\setcounter{subsection}{0}
\setcounter{subsubsection}{0}
\setcounter{equation}{0}
\setcounter{figure}{0}
\setcounter{table}{0}
\renewcommand{\thesection}{S\arabic{section}}
\renewcommand{\thesubsection}{\thesection.\arabic{subsection}}
\renewcommand{\thefigure}{S\arabic{figure}}
\renewcommand{\thetable}{S\arabic{table}}
\renewcommand{\theHsection}{Supplementary.\arabic{section}}
\renewcommand{\theHsubsection}{Supplementary.\arabic{section}.\arabic{subsection}}
\renewcommand{\theHequation}{Supplementary.\arabic{section}.\arabic{equation}}
\renewcommand{\theHfigure}{Supplementary.\arabic{figure}}
\renewcommand{\theHtable}{Supplementary.\arabic{table}}
\renewcommand{\refname}{Supplementary References}
\begin{center}
\vspace*{0.7em}
{\LARGE\bfseries Supplementary Information}\par\vspace{0.65em}
{\large Contextual Probability and Operational Minimality in Tug of War Decision Making}\par\vspace{0.65em}
{\normalsize Song-Ju Kim}\par\vspace{0.5em}
{\normalsize 14 October 2026}\par\vspace{1.2em}
\end{center}

\noindent These supplementary analyses provide classical control constructions, an alternative semantic information cost, a calibrated preparation-noncontextuality test, additional calibration families, and technically distinct QTOW realizations. The recovery--reset instrument in the main manuscript and the weak-readout/rotation instruments in Sections S5--S10 have different update laws. Parameters and witness values from one instrument are not transferred to another.\par\medskip

\section{Optional classical TOW learner with context-labelled C5 diagnostics}
\label{SI:app:classical-tow-c5-example}

\noindent\textit{Purpose and scope.}
This appendix preserves a constructive example in which a fully classical,
reward-driven two-state TOW learner is equipped with an \emph{independently
specified} five-context diagnostic service. Its value is to exhibit a
precisely quantified, learning-controlled transition in the \emph{diagnostic}
statistics from globally noncontextual-extendible to nonextendible
five-event tables without assuming quantum probability. The diagnostic
contexts and their absolute response calibration are additional
specifications; neither follows from native TOW learning, L1--L4, or the
conservation constraint. The example therefore does \emph{not} show that
native TOW learning itself requires nonclassical probability or that
all classical stochastic simulations fail. A matched noncontextual
control is included to make this independence explicit.

\subsection{Classical learner plus an independently specified diagnostic}
\label{SI:subsec:classical-learning-contextuality}

The foregoing distinction can be made explicit without quantum premises.
Attach to the classical learner of Eq.~(\ref*{eq:cf-native-kernel})
five classical context-local diagnostic responses. Set
\begin{equation}
 d=(-2,-1,0,1,2),\quad \epsilon=\frac1{2000},\qquad
 p_i(+1)=\frac49+\epsilon d_i,\quad
 p_i(-1)=\frac13+\epsilon d_i .
 \label{SI:eq:cf-classical-diagnostic}
\end{equation}
The conditional joint distribution for each adjacent pair is exactly
Eq.~(\ref*{eq:cf-c5-tables}), evaluated at the given value of $h$.
All probabilities are positive and normalized. Each event $e_i$ has the
same Yes marginal in its two contexts, although the joint sampling rule
must be supplied with a context label. If the learner's ensemble bias is $m$,
\begin{equation}
 p_i(m)=\frac{7+m}{18}+\epsilon d_i,\qquad
 K(m)=\frac{35+5m}{18},\qquad
 K(m)>2\ \Longleftrightarrow\ m>\frac15.
 \label{SI:eq:cf-learning-K}
\end{equation}
For $P_A=0.8$, $P_B=0.2$, $\chi=0.6$, and $\gamma=0.1$,
Eq.~(\ref*{eq:cf-native-learning-gain}) gives $m_*=0.6$ and
$K(m_*)=19/9>2$, whereas the uninformed $m_0=0$ preparation yields
$K(m_0)=35/18<2$. This is a fully classical, reward-conditioned
\emph{construction of locally classical but globally nonextendible
diagnostic probabilities}. All finite interleavings of these diagnostics
and native reward learning remain ordinary positive Markov processes
when the selected context is retained as a classical input.

\emph{Independence control.} The learning mechanism and the response
contrast need not select the contextual absolute calibration.
Indeed, replacing Eq.~\eqref{SI:eq:cf-classical-diagnostic} by
\begin{equation}
 p_i^{\rm NC}(-1)=\frac14+\epsilon d_i,\qquad
 p_i^{\rm NC}(+1)=\frac{13}{36}+\epsilon d_i
 \label{SI:eq:cf-classical-control}
\end{equation}
preserves the same learning process and the same sensitivity
$p_i(+1)-p_i(-1)=1/9$, but yields $K\le65/36<2$
for every mixture of the two hidden states. These alternative
diagnostics admit a common noncontextual joint model.
Thus, the classical-first result demonstrates a \emph{mathematical
connection and a reachable example}, not the inevitable onset of
contextuality from L1--L4. The five-event diagnostic and its
absolute response calibration are additional operational tasks,
whose costs and physical justification must be reported.

\noindent\textbf{Interpretation.}
The obstruction arises when \emph{context-locally classical} statistics
are required to share a single predetermined, context-independent
outcome-value semantics. It does not mean that the stochastic TOW
learner ceases to be classical or that no classical sequential
simulation exists. Its significance for operational minimality is
that a shared five-event description has an exactly measurable
repair cost once the chosen diagnostic service enters the regime
$K>2$.
\section{Classical information needed by a restricted common-event simulator}\label{SI:app:information-cost}
Assume the five contexts $C\in\{0,\ldots,4\}$ are sampled uniformly and independently of a preparatory variable $\Lambda$. A semantics-preserving competitor is allowed an additional context-correlated classical label $M$, but it must form one global stable-set valuation $V=V(\Lambda,M)\in\{0,1\}^5$, with $V_i V_{i+1}=0$, and report only $(V_C,V_{C+1})$ on the requested edge. In this \emph{restricted} completion class define $J_{\rm sem}=\inf I(C;M\mid\Lambda)$ over models reproducing the prescribed context-local tables. This is not an unrestricted classical memory or information cost.

For any such model reproducing a consistent exclusive $C_5$ box with $K>2$, let $Z=V_C+V_{C+1}\in\{0,1\}$. Under the actual law $P(C,\Lambda,M)$, the five shared marginals imply $P(Z=1)=2K/5$. Compare with the law $Q$ that keeps the actual $(\Lambda,M)$ marginal but resamples $C$ independently and uniformly. Under $Q$ each stable-set valuation contains at most two Yes vertices; thus $Q(Z=1)\le4/5$. Since $C\perp\Lambda$ under $P$, the chain rule gives $D_{\rm KL}(P_{C\Lambda M}\|Q_{C\Lambda M})=I_P(C;M\mid\Lambda)$. Data processing through $Z$ and monotonicity of binary relative entropy in its second argument for $2K/5>4/5$ yield
\[
 J_{\rm sem}\ge D_{\rm Ber}\!\left(\frac{2K}{5}\middle\|\frac45\right),\qquad
 D_{\rm Ber}(p\|q)=p\log_2\frac pq+(1-p)\log_2\frac{1-p}{1-q}.
\]
For $K\le2$ the general bound used here is zero. In the symmetric case, put $p=2K/5$ and select one of the five size-two independent sets conditional on the measured edge: the unique set avoiding that edge with probability $1-p$, and each of the other four with probability $p/4$. The resulting relative edge output probabilities are $(p/2,p/2,1-p)$ in every context, while the unconditional five-valuation distribution is uniform. Direct evaluation gives $I(C;V)=D_{\rm Ber}(p\|4/5)$, attaining the lower bound in this class.

For ideal quantum-compatible symmetric KCBS $K=\sqrt5$ the exact cost is $D_{\rm Ber}(2/\sqrt5\|4/5)\approx0.04665765$ bits. For the optional classical learner--diagnostic box in Supplementary Sec.~\ref{SI:app:classical-tow-c5-example} with $m=0.6$, $K=19/9$, the theorem implies $J_{\rm sem}\ge0.009469$ bits. Its nonuniform diagnostic response means the tightness construction above cannot be asserted for that specific learned table merely from its $K$ value. The label $M$ is a classical contextual bookkeeping register, not the qutrit auxiliary mode $|\perp\rangle$.
\section{A reference-calibrated preparation-noncontextuality test}\label{SI:si:preparation-noncontextuality}

This section considers a second contextuality criterion for the recovery--reset instrument of the main manuscript. In contrast to the occurrence-mismatch criterion of main-manuscript Eq.~(\ref*{eq:unified-finite-mismatch}), it tests \emph{preparation noncontextuality}: operationally equivalent preparation procedures must have identical ontic probability distributions \cite{SI:Spekkens2005}. The assumption concerns preparation distributions, not deterministic outcomes or context-independent measurement response functions. The finite construction uses the same stationary, unselected learned source $P$ represented by $\rho_m=\bar\rho_*$, but also requires independently callable positive recovery-gate words, feedforward, repeated source preparation, classical mixing, and suspension of bandit reward updates during diagnostics.

\subsection{Reference preparations and observable calibration errors}
For each of the five three-outcome contexts $C_i=(e_i,e_{i+1})$, let $L_i$ denote the first Yes outcome in context $C_i$ and $R_i$ the second Yes outcome in context $C_{i-1}$ (indices modulo five). Their actual effects may differ. They are two occurrences of the nominal event $e_i$, and in context $C_i$ the outcomes $L_i$ and $R_{i+1}$ are exclusive. Define
\begin{equation}\label{SI:eq:si-pnc-k}
 K_{\mathrm{av}}(P)=\frac12\sum_{i=0}^{4}\bigl[\Pr(L_i\mid P)+\Pr(R_i\mid P)\bigr].
\end{equation}
Suppose one reference preparation $P_{\mathrm{ref}}$ admits the exact \emph{operational} mixture equivalences
\begin{equation}\label{SI:eq:si-pnc-preps}
 P_{\mathrm{ref}}\simeq \alpha P+(1-\alpha)P_{\mathrm{comp}}
 \simeq a_iP_{i,+}+(1-a_i)P_{i,-},\qquad i\in\mathbb Z_5,
\end{equation}
where $0<\alpha,a_i<1$, and $\simeq$ means identical probabilities for every allowed continuation of the preparation. The same pair $P_{i,+},P_{i,-}$ must be tested against \emph{both} occurrences of event $i$. For $c\in\{L,R\}$, let
\begin{equation}\label{SI:eq:si-pnc-error}
 \epsilon_{i,+}^{c}=\Pr(\mathrm{No}\mid P_{i,+},c),\qquad
 \epsilon_{i,-}^{c}=\Pr(\mathrm{Yes}\mid P_{i,-},c),\qquad
 E=\sum_{i,c}\bigl[a_i\epsilon_{i,+}^{c}+(1-a_i)\epsilon_{i,-}^{c}\bigr].
\end{equation}
The errors are experimentally meaningful conditional probabilities; neither equality of nominal measurement effects nor perfect calibration is assumed.

\subsection{Classical bound under preparation noncontextuality}
\noindent\textbf{Proposition.} \emph{Any classical ontological model (of arbitrary finite or measurable ontic state space) reproducing the preparations, classical mixtures and exclusive three-outcome context responses of Eqs.~\eqref{SI:eq:si-pnc-preps}--\eqref{SI:eq:si-pnc-error}, and satisfying preparation noncontextuality, obeys}
\begin{equation}\label{SI:eq:recovery-PNC-bound}
 \boxed{K_{\mathrm{av}}(P)\le 2+\frac{3E}{2\alpha}.}
\end{equation}
\noindent\textit{Proof.} Write $\mu$ for ontic preparation measures. Operational equivalence and preparation noncontextuality imply a \emph{single} reference measure
$\mu_{\mathrm{ref}}=\alpha\mu_P+(1-\alpha)\mu_{\mathrm{comp}}=a_i\mu_{i,+}+(1-a_i)\mu_{i,-}$.
On $\mu_{\mathrm{ref}}$, define a hypothetical common binary calibration label $H_i$ by drawing from the posterior probability $a_i\,d\mu_{i,+}/d\mu_{\mathrm{ref}}$; use independent conditional draws for the five $i$. Conditionally on the ontic state, draw each physical three-outcome response from its actual stochastic response function. The coupling is solely a proof device; the model need not posit deterministic observed event values. For either occurrence $c$ of event $i$, let $Y_i^c$ be its binary Yes indicator. Direct conditioning on the calibration label gives
\[
 \Pr(H_i\ne Y_i^c\mid P_{\mathrm{ref}})=
 a_i\epsilon_{i,+}^c+(1-a_i)\epsilon_{i,-}^c.
\]
Because $\mu_{\mathrm{ref}}\ge\alpha\mu_P$ as measures, the corresponding probability on $P$ is at most $1/\alpha$ times this reference mismatch. For any joint assignment of the five mathematical labels and the five exclusive context outcomes, the odd-cycle combinatorial inequality and exclusivity give
\[
 \sum_i H_i\le 2+\sum_iH_iH_{i+1},\qquad
 H_iH_{i+1}\le |H_i-L_i|+|H_{i+1}-R_{i+1}|.
\]
Moreover, $\frac12\sum_i(L_i+R_i)\le \sum_iH_i+\frac12\sum_i(|H_i-L_i|+|H_i-R_i|)$. Combining them yields the pointwise bound
\[
 \frac12\sum_i(L_i+R_i)\le 2+\frac32\sum_{i,c}|H_i-Y_i^c|.
\]
Taking expectation in the coupled target experiment and applying the mismatch bounds proves Eq.~\eqref{SI:eq:recovery-PNC-bound}. $\square$

\subsection{Finite positive-word realization and independent verification}
Take the recovered binary readout $K_k=W_kM_k$, the learned mixture $\rho_m=0.8|\ell_+\rangle\langle\ell_+|+0.2|\ell_-\rangle\langle\ell_-|$, and the seven finite positive words in main-manuscript Appendix~\ref*{app:modular-recovery}. For each context, the two successive readouts produce the four raw outcomes $AA,BB,AB,BA$; the reported three outcomes are $AA$, $BB$, and $AB\lor BA$. For each raw outcome, additional positive recovery words are sampled conditionally, with their probabilities chosen so that the outcome-forgotten context instrument maps $I_3/3$ \emph{exactly} to $I_3/3$ in the ideal real-number control model. The selected supports contain nine positive weights per context, obtained from a nonsingular $9\times9$ linear system imposing the four branch normalizations and five independent traceless matrix constraints. The two calibration preparations $P_{i,+}$ and $P_{i,-}$ are the $AA$-postselected and complementary-output conditional states of the same corrected context instrument acting on $I_3/3$. Thus their mixture equals $I_3/3$, with $a_i=\Pr(AA\mid I_3/3,C_i)$, by construction.

Six finite positive words of lengths $0,92,88,81,73,21$ applied to independent copies of $\rho_m$ produce an exact common reference $\rho_{\mathrm{ref}}=I_3/3$ with positive weights $w_j$, the unique solution of a $6\times6$ system imposing trace normalization and five independent symmetric traceless entries. The identity word has $w_0>1/3$. The \emph{same} six states, with weights $v_0=(3w_0-1)/2$ and $v_j=3w_j/2$ for $j>0$, prepare $\rho_{\mathrm{comp}}=(I_3-\rho_m)/2$. Consequently
\[
 \frac{I_3}{3}=\frac13\rho_m+\frac23\rho_{\mathrm{comp}}
 =a_i\rho_{i,+}+(1-a_i)\rho_{i,-}
\]
for each $i$, with $\alpha=1/3$. Positivity of all the exact-real linear-system solutions is certified by interval arithmetic. This construction assumes idealized high-precision classical mixture controls; it is not a finite-digit or experimentally demonstrated equivalence test.

\begin{table}[t]
\centering\small
\caption{Finite positive-word calibrated preparation-noncontextuality witness for the unselected learned source at $m=0.6$. The error $E$ is defined in Eq.~\eqref{SI:eq:si-pnc-error}.}
\label{SI:tab:si-pnc-numerics}
\begin{tabular}{@{}ll@{}}\toprule
Quantity & Value \\\midrule
$K_{\mathrm{av}}$ & $2.0676577898997393$\\
$E$ & $0.0003768918497872960$\\
$2+3E/(2\alpha)$, $\alpha=1/3$ & $2.0016960133240428$\\
Excess over preparation-noncontextual bound & $0.0659617765756965$\\
Smallest conditional correction weight & $0.00233121530637$\\
Smallest six-word source weight & $0.0341460296793$\\
Largest individual cross-calibration error & $0.00010859889431$\\\bottomrule
\end{tabular}
\end{table}
Table~\ref{SI:tab:si-pnc-numerics} provides the distinct calibrated witness; it is not the CbD-type excess of $0.058309\ldots$ in the main manuscript. The accompanying file \texttt{verification/pnc\_verified\_manifest.json} specifies the positive words, conditional branches, and mixture supports. The independent programs \texttt{verify\_pnc\_high\_precision.py} and \texttt{verify\_pnc\_interval.py} reconstruct the weights from the displayed finite supports (without rerunning their search), verify the five mixture identities, and certify a strictly positive excess greater than $0.0659$. The archived unrounded source manifest is also supplied for direct execution. No classical model obeying the stated preparation-noncontextuality assumption can reproduce these idealized operational equivalences and statistics; preparation-contextual classical simulators are not excluded. In a physical test, finite precision and finite data require separate, explicitly validated operational-equivalence and noise analyses \cite{SI:Mazurek2016}.
\section{Alternative polynomial calibration family and finite-precision limitations}\label{SI:si:polynomial-calibration}
\subsection*{Polynomial comparison}

\subsection*{Conditional rigidity at the maximal point}

For comparison, a second independently specified calibration family illustrates whether the
\emph{probability vector} at $K=\sqrt5$ can be selected by
calibration and composition, without presupposing a Hilbert-space
model. Put $t=1/\sqrt5$ and consider the one-parameter
context-local calibration family
\begin{equation}
 q_i(\vartheta)=t-2\vartheta^2+\vartheta v_i,\quad
 v=\left(2,-\frac12,-\frac12,-\frac12,-\frac12\right),\quad
 0\le\vartheta\le\frac1{100}.
 \label{SI:eq:appcf-cal-family}
\end{equation}
For this stated parameter range, these are nonnegative exclusive
local probability tables. They can be sampled by a classical
controller given an experimental context and the control
$\vartheta$. The existence of this calibrated \emph{continuous
operation family} is an additional resource assumption; it is
not deduced from native reward updates. In particular, the value
$t=1/\sqrt5$ is built into the chosen family, so the result
below is a conditional rigidity theorem, not a parameter-free
reconstruction of quantum theory.

\noindent\textbf{Proposition (polynomial-calibration rigidity at the maximal face).}
Assume that a preparation set contains the family in
Eq.~\eqref{SI:eq:appcf-cal-family}, is closed under classical mixing,
and obeys mixed-preparation GE2 between \emph{every} pair of its
accessible preparations. If an accessible preparation $p$ has
$\sum_i p_i=\sqrt5$, then
\begin{equation}
 \boxed{\displaystyle p_0=p_1=\cdots=p_4=\frac1{\sqrt5}.}
 \label{SI:eq:appcf-rigidity}
\end{equation}
\emph{Proof.} For each $s$, mixed-preparation GE2 implies
\[
 \sum_i p_i q_{2i+s}(\vartheta)
 =1-2\sqrt5\,\vartheta^2
  +\frac52\vartheta\bigl(p_{i_s}-t\bigr)\le1,
\]
where $2i_s+s\equiv0\pmod5$.
Thus $p_{i_s}-t\le(4/\sqrt5)\vartheta$.
Since the condition holds for every arbitrarily small positive
$\vartheta$, we obtain $p_i\le t$ for all $i$.
Their sum is $5t$, forcing equality throughout.
\hfill$\square$

The family \eqref{SI:eq:appcf-cal-family} has itself been checked to be
GE2-compatible, including classical mixtures with the
noncontextual stable-set preparations, so these premises
do not require inconsistent probability tables.
In contrast, a \emph{finite} menu of nonzero calibration settings
only bounds the deviation from Eq.~\eqref{SI:eq:appcf-rigidity};
it does not enforce exact equality. The bound and its
finite-resolution limitations are given in the Appendix.

\subsection{Calibration family and precision limitations}

The alternative polynomial family of Eq.~\eqref{SI:eq:appcf-cal-family} is an
\emph{assumed accessible} operation. For
$0\le\vartheta,\varphi\le1/100$ the pairwise
GE2 inequalities for $q(\vartheta)$ and $q(\varphi)$
can be checked by partitioning the product exclusivity
graph into its cliques of sizes at most five.
The extremal size-five cliques are the permutations
$(i,2i+s)$ and their transposes. The former have
\[
 \sum_i q_i(\vartheta)q_{2i+s}(\varphi)
 =5(t-2\vartheta^2)(t-2\varphi^2)
 +\vartheta\varphi S_s,\qquad
 S_0=5,\ \ S_{s\ne0}=-\frac54 .
\]
The quadratic deficits dominate the mixed term throughout
the specified interval; the smaller cliques also satisfy GE2.
As a separate finite-setting limitation, if a
maximal preparation $p$ obeys the five mixed GE2 inequalities
with an additive tolerance $\delta\ge0$
at a given $\vartheta>0$, then
\[
 \max_i(p_i-t)\le
 \frac4{\sqrt5}\vartheta+\frac{2\delta}{5\vartheta}.
\]
A finite nonzero set of settings therefore does not alone
force exact equality at the maximal face.

These constraints refer to product \emph{Yes events} of two
copies. They do not claim an experimentally implemented
complete joint measurement instrument for every compound
outcome, all-copy global exclusivity, or a derivation of a
quantum probability theory.
\section{Weak-Readout and Unitary-Feedback QTOW (Alternative Model)}\label{SI:secA1}

This section studies a weak-readout instrument followed by reward-conditioned unitary feedback. Its update law and fixed-angle projectors differ from the recovery--reset instrument in the main manuscript, Secs.~\ref*{sec:qtow-implementation} and~\ref*{sec:qtow-reachability}. The two constructions share a two-arm interface but do not share the same reward channel.

\subsection{Operational sequence of one trial}
One trial consists of the following operations:
(i) an \emph{ordinary decision instrument} $\{K_A,K_B\}$ acts on the qutrit, yielding $d_t\in\{A,B\}$ with probabilities $p(d_t|\psi_t)=\langle\psi_t|E_{d_t}|\psi_t\rangle$ and an outcome-conditioned post-decision state $|\psi_t^{(d_t)}\rangle$;
(ii) an \emph{external reward} $r_t\in\{\mathrm{win},\mathrm{lose}\}$ is generated by the environment conditional on $d_t$;
(iii) the internal state is updated by a decision- and reward-conditioned unitary $U_{d_t,r_t}$.
Optionally, a \emph{contextual probe} $P_i$ may be inserted immediately before the ordinary decision instrument, defining a different measurement context.

The ordinary sequence above defines the physical time ordering of the QTOW learning process. The optional probing and KCBS constructions developed below are not claims that all probe contexts occur within that ordinary timeline; rather, they provide counterfactual and logical tests of whether the extended operation family admits a non-contextual classical representation.

We distinguish non-contextual outcome assignments, context-labelled classical sampling, and the stipulated quantum realization in Secs.~\ref*{sec:classical-first}--\ref*{sec:qtow-implementation}; these descriptions impose different assumptions.

\subsubsection*{Lemma (Counterfactual diagnostic: a non-commuting probe changes decision statistics)}
\label{SI:lem:A1}

The lemma concerns a counterfactual probe-insertion protocol rather than the ordinary QTOW time ordering. It shows how a non-commuting intervention on the \emph{same} internal state can change the statistics of the subsequent generalized decision instrument even when the pre-probe state is identical.

\noindent\textit{Statement.}
Let the ordinary decision be described by the effects $E_A,E_B$ defined below, with $0<\eta<1$, and insert immediately before it a binary projective probe $P$ that does not commute with $E_A$. Then, for generic internal states, the probability of deciding $A$ differs between the ``decision-only'' and ``probe-then-decision'' protocols.

\noindent\textit{Setup.}
Let
\begin{equation}
|\psi\rangle=a|A\rangle+b|B\rangle+c|\perp\rangle,
\qquad |a|^2+|b|^2+|c|^2=1,
\end{equation}
and let
\begin{align}
E_A &= \frac{1+\eta}{2}|A\rangle\langle A|+\frac{1-\eta}{2}|B\rangle\langle B|+\frac{1}{2}|\perp\rangle\langle\perp|,\\
E_B &= I-E_A.
\end{align}
The decision-only probability is
\begin{equation}
P_{\mathrm{dec}}(A)=\langle\psi|E_A|\psi\rangle.
\end{equation}

Consider the rank-1 probe
\begin{equation}
P(\beta)=|v(\beta)\rangle\langle v(\beta)|,\qquad
|v(\beta)\rangle=\cos\beta\,|A\rangle+\sin\beta\,|\perp\rangle,
\qquad \beta\in(0,\pi/2),
\end{equation}
for which $[E_A,P(\beta)]\neq0$ whenever $\eta>0$. With $\rho=|\psi\rangle\langle\psi|$, the unrecorded probe produces the L\"uders state
\begin{equation}
\rho'=P(\beta)\rho P(\beta)+(I-P(\beta))\rho(I-P(\beta)),
\end{equation}
and the subsequent decision probability is
\begin{equation}
P_{P\rightarrow\mathrm{dec}}(A)=\mathrm{Tr}(E_A\rho').
\end{equation}

\noindent\textit{Explicit witness of context dependence.}
Take the same initial internal state $|\psi\rangle=|\perp\rangle$. Without the probe,
\begin{equation}
P_{\mathrm{dec}}(A)=\frac{1}{2}.
\end{equation}
After the probe, the population transferred into $|A\rangle$ is $2\sin^2\beta\cos^2\beta$, and one obtains
\begin{equation}
P_{P\rightarrow\mathrm{dec}}(A)
=\frac{1}{2}+\eta\sin^2\beta\cos^2\beta
=\frac{1}{2}+\frac{\eta}{4}\sin^2(2\beta)
>\frac{1}{2}.
\end{equation}
Hence
\begin{equation}
P_{P\rightarrow\mathrm{dec}}(A)\neq P_{\mathrm{dec}}(A),
\end{equation}
even though the internal state before the optional probe is identical.

\noindent\textit{Conclusion.}
A non-commuting intermediate probe changes subsequent decision statistics by measurement disturbance. This order effect alone is not a Kochen--Specker contextuality witness. The latter requires the distinct, compatible, exclusive joint diagnostic contexts specified below. Such probes are not part of every bandit decision trial. In the remainder of this supplementary section, the KCBS construction \cite{SI:KCBS2008PRL} is used as a logical diagnostic to show that an appropriate admissible qutrit probe family cannot be jointly embedded into a non-contextual hidden-variable model.

\subsection*{Classical TOW as a constrained state-update process}

For a two-option Tug-of-War (TOW) decision-making model, the coupled preference state may be represented by a two-dimensional real vector
\begin{equation}
\mathbf{x}_t = (x_A(t), x_B(t)) \in \mathbb{R}^2,
\end{equation}
which evolves under the conservation law
\begin{equation}
x_A(t) + x_B(t) = \mathrm{const}.
\end{equation}

Due to this constraint, the effective number of degrees of freedom is reduced to one.
The state update rule can be written as
\begin{equation}
\mathbf{x}_{t+1} = F_{k_t,r_t}(\mathbf{x}_t) + \boldsymbol{\xi}_t,
\end{equation}
where the update function $F_{k_t,r_t}$ depends on the selected arm $k_t\in\{A,B\}$ and reward outcome $r_t$, and $\boldsymbol{\xi}_t$ represents stochastic fluctuations, subject to preservation of the coupled-resource constraint.

In this classical formulation, the stochastic update can be represented by an ordinary state-space transition kernel. The conservation constraint alone imposes no quantum measurement rule.

\subsection*{Quantum representation of the internal state}

\subsubsection*{Internal state as a Hilbert-space vector}

In the quantum extension of TOW, the internal state is represented as a normalized vector
\begin{equation}
|\psi_t\rangle \in \mathcal{H}, \qquad \dim \mathcal{H} = 3.
\end{equation}

The Hilbert-space basis is chosen such that:
\begin{itemize}
\item $|A\rangle$ and $|B\rangle$ span the two-dimensional decision subspace corresponding to the two competing options,
\item $|\perp\rangle$ is an auxiliary state associated with contextual probing.
\end{itemize}

The qutrit is the minimal dimensional realization of the specific KCBS-based construction used here. A two-dimensional Hilbert space is insufficient for the standard single-system KCBS projective contextuality scenario \cite{SI:KCBS2008PRL}. This dimensional statement does not imply that every quantum-like learning architecture requires a qutrit.

\subsubsection*{Conservation law and decision-conditioned unitary feedback}

For this chosen quantum realization, the state is normalized:
\begin{equation}
\langle \psi_t | \psi_t \rangle = 1.
\end{equation}

Within a closed Hilbert-space realization, the feedback step following a decision and reward is taken to be unitary:
\begin{equation}
|\psi_{t+1}\rangle = U_{d_t,r_t}|\psi_t^{(d_t)}\rangle,
\qquad U_{d_t,r_t}\in\mathrm{U}(3).
\end{equation}
The signed rotations chosen for the real-amplitude decision plane move the conditional preference toward the selected arm after a win and away from it after a loss. This directional update does not by itself guarantee superior stationary decision performance. The auxiliary state $|\perp\rangle$ is left invariant by these ordinary feedback rotations and supplies the third mode used by the qutrit probe construction.

\subsection*{Ordinary decisions and contextual probes}

\subsubsection*{Generalized decision instrument}

The ordinary decision is described by the two effects
\begin{align}
E_A &= \frac{1+\eta}{2}|A\rangle\langle A|+\frac{1-\eta}{2}|B\rangle\langle B|+\frac{1}{2}|\perp\rangle\langle\perp|,\\
E_B &= I-E_A,
\end{align}
with $0<\eta<1$, together with the L\"uders Kraus operators
\begin{equation}
K_A=\sqrt{E_A},\qquad K_B=\sqrt{E_B}.
\end{equation}
The probability of choosing option $d\in\{A,B\}$ is
\begin{equation}
P(d|\psi_t)=\langle\psi_t|E_d|\psi_t\rangle,
\end{equation}
and the corresponding conditional state is
\begin{equation}
|\psi_t^{(d)}\rangle=
\frac{K_d|\psi_t\rangle}{\sqrt{P(d|\psi_t)}}.
\end{equation}
For finite sharpness the decision changes the state but does not erase all pre-decision amplitude information. On the ordinary two-dimensional learning subspace, the projective $A/B$ readout is recovered as the limit $\eta\to1$.

\subsubsection*{Contextual probing}

Contextual probes are represented by a distinct set of projective measurements
\begin{equation}
P_i = |v_i\rangle\langle v_i|,
\qquad i = 1, \dots, 5.
\end{equation}

In general, these projectors satisfy
\begin{equation}
[E_A, P_i] \neq 0,
\qquad
[P_i, P_j] \neq 0
\quad \text{for } j\not\equiv i, i\pm1\pmod{5}.
\end{equation}

Hence, the ordinary decision effects and contextual probes, as well as non-adjacent probes among themselves, can be mutually incompatible.

In what follows, the KCBS scenario is defined by the operational compatibility requirement that only adjacent pairs $(P_i,P_{i+1})$ admit a joint measurement (e.g., via a three-outcome projective measurement $\{P_i,\,P_{i+1},\,I-P_i-P_{i+1}\}$), while non-adjacent pairs do not share any joint measurement in the chosen probe set.

\subsection*{KCBS structure in the QTOW state space}

\subsubsection*{KCBS projectors}

Within the qutrit Hilbert space, there exists a set of five projectors $\{P_i\}$ satisfying the KCBS compatibility relations \cite{SI:KCBS2008PRL}:
\begin{itemize}
\item adjacent pairs satisfy $P_iP_{i+1}=0$ and are jointly measurable, e.g., by a single measurement with effects $\{P_i,\ P_{i+1},\ I-P_i-P_{i+1}\}$,
\item the remaining pairs are taken as incompatible in the chosen KCBS compatibility structure.
\end{itemize}

For any non-contextual hidden-variable theory, the following inequality holds:
\begin{equation}
\sum_{i=1}^{5} \langle P_i \rangle \le 2.
\end{equation}

\subsubsection*{Quantum prediction}

Before introducing the KCBS construction, we emphasize that it does not represent the physical time ordering of the QTOW process. Rather, it provides a logical test of whether the specified qutrit probe family can be jointly embedded into a non-contextual hidden-variable model.

For a given quantum state $|\psi\rangle$,
\begin{equation}
\langle P_i \rangle = \langle \psi | P_i | \psi \rangle.
\end{equation}

It is a standard result for the KCBS exclusivity (5-cycle) scenario on a qutrit that the maximal quantum value of the projector sum equals $\sqrt{5}$.

For rank-1 projectors $\{P_i\}_{i=1}^{5}$ whose compatibility (exclusivity) relations form a KCBS $5$-cycle (adjacent pairs exclusive/compatible), quantum mechanics allows
\begin{equation}
\max_{|\psi\rangle}\ \sum_{i=1}^{5} \langle P_i \rangle = \sqrt{5},
\end{equation}
so there exist states $|\psi\rangle$ for which
\begin{equation}
\sum_{i=1}^{5} \langle P_i \rangle > 2.
\label{SI:eq:kcbs-violation-app}
\end{equation}

Thus, violations of the KCBS non-contextuality inequality are permitted within the QTOW framework.

\subsection*{Failure of context-independent outcome-value assignments}

Suppose one assumes an underlying classical hidden variable $\lambda$.
Each projector $P_i$ is then assigned a definite value
\begin{equation}
P_i(\lambda) \in \{0,1\}.
\end{equation}

Non-contextuality requires that
\begin{equation}
P_i(\lambda) \text{ is uniquely defined, independent of context}.
\end{equation}
In the KCBS setting, each $P_i$ appears in two different compatible contexts $(P_{i-1},P_i)$ and $(P_i,P_{i+1})$; non-contextuality asserts that the predetermined value assigned to $P_i$ is the same in both contexts.

Together with the essential pointwise adjacent-exclusivity condition $P_i(\lambda)P_{i+1}(\lambda)=0$ (indices modulo five), these assumptions imply the constraint
\begin{equation}
\sum_{i=1}^{5} P_i(\lambda) \le 2,
\end{equation}
which contradicts the quantum prediction in Eq.~\eqref{SI:eq:kcbs-violation-app}.

Therefore, a non-contextual hidden-variable assignment of pre-existing values is inconsistent with the statistics of the specified qutrit state under the KCBS probe family.

\subsubsection*{Auxiliary-axis KCBS pentagon}
For completeness, we give a KCBS pentagon whose symmetry axis is the auxiliary mode $|\perp\rangle$. This geometry attains the maximum $\sqrt5$ on the auxiliary state, whereas the reward-learning trajectory is confined to the $A/B$ plane. The alternative finite-time dynamical result uses a distinct, fixed orientation with its symmetry axis along $|A\rangle$ (Supplementary Sec.~\ref{SI:app:alt-rotation-reachability}).

Let $\alpha$ satisfy
\begin{equation}
\cos^2\alpha=\frac{1}{\sqrt{5}},\qquad \sin^2\alpha=1-\frac{1}{\sqrt{5}}.
\end{equation}
Define five unit vectors $|v_i\rangle\in\mathbb{R}^3$ by
\begin{equation}
|v_i\rangle=
\begin{pmatrix}
\sin\alpha\cos\phi_i\\
\sin\alpha\sin\phi_i\\
\cos\alpha
\end{pmatrix},
\qquad
\phi_i=\frac{4\pi i}{5},\qquad i=1,\dots,5.
\end{equation}
One verifies that adjacent vectors are orthogonal,
\begin{equation}
\langle v_i|v_{i+1}\rangle = 0\qquad (\text{indices mod }5),
\end{equation}
so adjacent projectors $P_i:=|v_i\rangle\langle v_i|$ are exclusive/compatible in the KCBS cycle.
(Operationally, the joint measurability of an adjacent pair can be implemented by a single three-outcome measurement with effects $\{P_i,\ P_{i+1},\ I-P_i-P_{i+1}\}$.)

Choose the state
\begin{equation}
|\psi\rangle = |z\rangle := (0,0,1)^T.
\end{equation}
Then for every $i$,
\begin{equation}
\langle P_i\rangle = |\langle v_i|z\rangle|^2 = \cos^2\alpha = \frac{1}{\sqrt{5}}.
\end{equation}
Hence the KCBS sum is
\begin{equation}
\sum_{i=1}^{5}\langle P_i\rangle = 5\cdot\frac{1}{\sqrt{5}}=\sqrt{5}>2,
\end{equation}
which violates the noncontextual bound and demonstrates KCBS contextuality for a single qutrit.

\subsubsection*{Role of the auxiliary degree of freedom in the present construction}

The auxiliary level $|\perp\rangle$ should be interpreted narrowly. It completes the qutrit used in the present QTOW realization, allowing the two-option decision subspace and the KCBS-compatible probing geometry to be represented within one finite state space. Its introduction is therefore a modeling choice tied to the construction analyzed here.

No universal claim is made that adaptive quantum learning requires a qutrit or that environmental estimates must be stored in a separate quantum level. Alternative architectures could use different Hilbert-space dimensions, mixed states, quantum channels, classical feedback, or enlarged classical state variables. The result established in this paper is instead conditional: once the qutrit state space and the admissible KCBS probe family are included, there exist state-and-probe configurations that violate the non-contextual bound.

This distinction is important because the KCBS witness diagnoses the representability of the chosen operation family; it does not derive the qutrit dimension from learning requirements. The auxiliary level therefore provides the minimal additional dimension required by the present KCBS-based construction, not a general lower bound on the memory required for decision making.

\subsection*{Operational scope}

The QTOW construction explicitly uses:
\begin{enumerate}
\item a finite-resource constraint represented in the closed Hilbert-space realization by norm preservation,
\item closed Hilbert-space updates represented by unitaries in the minimal realization,
\item a generalized decision instrument with state disturbance, together with projective contextual probing measurements.
\end{enumerate}

It does \emph{not} assume:
\begin{itemize}
\item quantum physical processes in the brain,
\item psychological axioms forcing quantum theory,
\item data fitting or parameter tuning,
\item a claim that sharp contextual diagnostics follow from native A/B reward learning alone; those diagnostics are added explicitly.
\end{itemize}

Within this explicit quantum-like realization, the KCBS probe family admits states that violate the non-contextual bound. A classical account of these statistics must give up at least one of the explicitly tested assumptions of a common outcome-deterministic noncontextual value assignment. A context-aware classical sampler remains possible.

\emph{Quantum Tug-of-War decision making should therefore be understood here as a compact single-state realization of a contextual operation family under stated representational constraints, rather than as an assumption-free derivation of quantum theory.}
\section{Asymmetric QTOW Feedback in General Reward Environments}\label{SI:secAppB}

\noindent\textit{Asymmetric-feedback setting.}
Define the external reward-probability sum (the environmental strength parameter)
\begin{equation}
g := P_A + P_B,
\end{equation}
where $P_A$ and $P_B$ are the environment's reward probabilities conditional on choosing arms $A$ and $B$, respectively.
Note that $g$ is not the per-trial reward probability, which is given by
$P(\mathrm{win\ at}\ t)=P_t(A)P_A+P_t(B)P_B$.

This section examines an asymmetric feedback law for the weak-readout, unitary-rotation model. Stationary reward improvement comparable to classical TOW is not established for this architecture.
The quantum extension considered here combines a persistent, state-disturbing decision process with norm-preserving feedback; the KCBS contextuality witness additionally depends on the qutrit state space and the compatibility geometry of the admissible projective probes.
At the same time, because the environmental strength parameter $g=P_A+P_B$ is generally unknown and must be estimated online, it is useful to record (briefly) that the same adaptive control of update asymmetry can also be implemented in a purely classical internal-state model.
We include this classical counterpart only as a reference baseline; it is not the focus of the present work.

The main finite-time example uses a symmetric update with $P_A+P_B=1$; the binary instrument itself and the fixed KCBS construction are defined for more general reward parameters. The equality $P_A+P_B=1$ is a choice of environment, not a probability normalization.
Rather, it is a convenient special case expressing a symmetry of the feedback magnitudes in the ordinary learning trajectory. The KCBS witness itself does not depend on this symmetry condition; it depends on the qutrit state and the chosen KCBS-compatible probe family.

One may propose an asymmetric-feedback generalization to
general environments in which $P_A + P_B \neq 1$.
In this appendix, we briefly outline such a generalization and clarify its relation to the
results for the recovery--reset instrument in the main manuscript.

\subsection*{Asymmetric update rule and generalized conservation}

The sum of the two arm reward probabilities
\begin{equation}
g = P_A + P_B
\end{equation}
satisfies $0\leq g\leq2$; the ratio below is finite only for $g<2$.
Asymmetric reward frequencies motivate unequal increments in a classical expected-update balance.

The classical TOW update rule is therefore generalized to asymmetric increments of the form
\begin{equation}
+1 \quad \text{(reward)}, \qquad -w \quad \text{(no reward)},
\end{equation}
where the asymmetry parameter $w$ is determined by the conservation condition as
\begin{equation}
w = \frac{g}{2-g}.
\end{equation}

In the classical baseline, this rule is motivated by balancing expected preference increments. It does not prove a corresponding stationary reward advantage for the quantum rotation rule.

\subsection*{Quantum implementation with asymmetric decision-conditioned feedback}

An asymmetric rule can be proposed in the qutrit feedback representation by retaining the decision-dependent sign structure of the weak-readout model of Supplementary Sec.~\ref{SI:secA1} while allowing different magnitudes for reward and no-reward feedback. For $0<g<2$, let $\theta>0$ denote the reward rotation magnitude and $\phi>0$ the no-reward magnitude. Then
\begin{align}
U_{A,\mathrm{win}} &= R_{AB}(-\theta)\oplus(1), &
U_{A,\mathrm{lose}} &= R_{AB}(+\phi)\oplus(1),\\
U_{B,\mathrm{win}} &= R_{AB}(+\theta)\oplus(1), &
U_{B,\mathrm{lose}} &= R_{AB}(-\phi)\oplus(1).
\end{align}
The asymmetry ratio is chosen as
\begin{equation}
\frac{\phi}{\theta}=w=\frac{g}{2-g}.
\end{equation}
Thus a reward always shifts the state toward the selected option, whereas a loss shifts it away; the environmental parameter $g$ controls only the relative feedback magnitude. Each feedback operation remains unitary, so norm preservation is maintained trial by trial. Near the boundaries of the real-amplitude chart, a state-dependent angle reduction requires the pure-state controller specified in Eq.~\eqref{SI:eq:clip-policy}; it does not automatically define a completely positive trace-preserving (CPTP) quantum channel on an unknown mixture.

The architecture therefore combines a partially state-preserving generalized decision instrument with unitary feedback. This distinction is separate from the KCBS witness, which still depends on the qutrit state and the compatibility geometry of the chosen projective probe family.

\subsection*{Online estimation of $g$ and adaptive asymmetry (quantum state + classical control)}

In realistic environments, $g=P_A+P_B$ is not known a priori.
A conservative and analytically transparent implementation is to keep an online estimate $\hat g_t$ as a \emph{classical} internal control parameter, while the decision state $|\psi_t\rangle$ remains a qutrit that undergoes outcome-conditioned generalized measurements followed by unitary feedback.

The adaptive QTOW feedback is applied to the post-decision state $|\psi_t^{(d_t)}\rangle$:
\begin{equation}
|\psi_{t+1}\rangle = U_{d_t,r_t}(\hat g_t)|\psi_t^{(d_t)}\rangle,
\qquad
\phi_t=w(\hat g_t)\theta,
\qquad
w(\hat g_t)=\frac{\hat g_t}{2-\hat g_t},
\end{equation}
with
\begin{align}
U_{A,\mathrm{win}} &= R_{AB}(-\theta)\oplus(1), &
U_{A,\mathrm{lose}} &= R_{AB}(+\phi_t)\oplus(1),\\
U_{B,\mathrm{win}} &= R_{AB}(+\theta)\oplus(1), &
U_{B,\mathrm{lose}} &= R_{AB}(-\phi_t)\oplus(1).
\end{align}
Each feedback step remains unitary; only the no-reward rotation magnitude adapts through the environmental estimate.

Because $g=P_A+P_B$ is not the reward probability of a single trial, it cannot in general be estimated consistently from the binary reward indicator alone without conditioning on the selected arm. Let $k_t\in\{A,B\}$ denote the selected arm and $I_t\in\{0,1\}$ the reward outcome. Here $\eta_t$ denotes a stochastic-approximation step size, distinct from the fixed decision sharpness $\eta$. A simple pair of stochastic estimators is
\begin{align}
\hat P_{A,t+1} &=
\begin{cases}
\hat P_{A,t}+\eta_t\!\left(I_t-\hat P_{A,t}\right), & k_t=A,\\
\hat P_{A,t}, & k_t=B,
\end{cases}\\
\hat P_{B,t+1} &=
\begin{cases}
\hat P_{B,t}, & k_t=A,\\
\hat P_{B,t}+\eta_t\!\left(I_t-\hat P_{B,t}\right), & k_t=B,
\end{cases}
\end{align}
with a suitable learning-rate sequence $\eta_t$. The environmental-strength estimate is then
\begin{equation}
\hat g_t=\Pi_{[\epsilon,\,2-\epsilon]}\!\left(\hat P_{A,t}+\hat P_{B,t}\right),
\end{equation}
where $0<\epsilon<1$ prevents $\hat g$ from approaching the singular endpoint $2$. Consistent estimation additionally requires that both arms continue to be sampled sufficiently often. This method separates roles cleanly: the qutrit represents the decision state in the present construction, while the classical control variables estimate the external reward probabilities.

\noindent\textit{Classical adaptive counterpart (baseline only).}
For comparison, one may also formulate an entirely classical adaptive TOW with a scalar preference variable $x_t$ (or a two-dimensional constrained state) updated by asymmetric increments $+1$ and $-w(\hat g_t)$, where the same online estimate $\hat g_t$ controls the asymmetry through $w(\hat g_t)=\hat g_t/(2-\hat g_t)$.
Such a classical formulation can reproduce history dependence, but it does not by itself imply Kochen--Specker contextuality. In the qutrit-based construction, the KCBS witness additionally relies on the qutrit state space and the compatibility geometry of the chosen probe family; non-commuting updates or probes alone do not guarantee a violation.
Accordingly, the classical baseline is recorded here only to clarify that the unknown-$g$ issue is not an obstacle unique to quantum implementations.

\noindent\textit{Asymptotic behavior.} These asymmetric angle choices do not establish a classical-TOW-matched stationary drift for arbitrary $g$. In particular, the fixed symmetric reward rotations at $g=1$ have an unbiased asymptotic ensemble in the contractive regime. Moreover, the formula $\phi/\theta=g/(2-g)$ inherited from the classical expected-increment argument is a \emph{candidate parametrization} in the quantum model, not an optimality theorem. A boundary rule depending on the hidden pure amplitude angle cannot be promoted without qualification to a CPTP map on an unspecified mixed preparation. These limitations do not change the exact interior one-trial result of Supplementary Sec.~\ref{SI:app:alt-rotation-reachability}.

\subsection*{Relation to contextuality results}

The generalized feedback rotation magnitudes do not by themselves determine whether a KCBS violation occurs. The KCBS witness depends on the qutrit state and on the compatibility geometry of the chosen probe family, not on the special symmetry condition $P_A+P_B=1$.

Accordingly, replacing the symmetric update by the asymmetric rule above does not, by itself, remove the availability of the KCBS probe family, provided that the same qutrit state space and admissible probes are retained. Whether a particular dynamical trajectory reaches a state that violates the KCBS bound is a separate state-preparation question. A detailed dynamical analysis outside the symmetric regime is left for future work.
\section{Optical-mode sketch for the weak-readout, unitary-feedback model}
\label{SI:secAppC}

\noindent\textit{Optical-mode implementation.}
We describe a possible optical-mode encoding of the weak-readout and unitary-feedback model of Supplementary Secs.~\ref{SI:secA1}--\ref{SI:secAppB}, including its fixed-angle contextual probes. It is not an implementation of the recovery--reset instrument and its independently scheduled recovery gates in the main manuscript.

No main-manuscript proof relies on the optical-mode sketch. The discussion establishes plausibility at the level of mode encoding and elementary transformations; it is not a complete experimental demonstration, and detailed calibration, compatibility verification, loss analysis, and state-preparation procedures remain to be specified.

For an explicit contextuality protocol, a non-commuting probe may be inserted before the ordinary generalized decision instrument. Such probe-inserted protocols define the contextual tests considered here and are distinct from the ordinary learning sequence shown in Fig.~\ref{SI:fig:qtow_concept}.

\subsection*{Photonic qutrit encoding}
A convenient qutrit can be encoded using optical path labels $u,d$ (upper and lower) and polarization labels $H,V$ (horizontal and vertical), selecting three modes:
\begin{equation}
|0\rangle \equiv |u,H\rangle,\quad
|1\rangle \equiv |u,V\rangle,\quad
|2\rangle \equiv |d,H\rangle,
\end{equation}
while the unused mode $|d,V\rangle$ is blocked or routed to a dump.
The decision subspace is identified as $\mathrm{span}\{|0\rangle,|1\rangle\}$, corresponding to options $A$ and $B$.

\subsection*{QTOW decision instrument and feedback update}
The ordinary $A/B$ decision can be implemented as a partial-strength polarization measurement: the polarization is weakly or partially coupled to a readout degree of freedom, producing two outcomes while retaining coherence in the conditional signal state. The measurement strength realizes the sharpness parameter $\eta$; the limit $\eta\to1$ approaches a projective polarization readout on the decision subspace.

After observing decision $d_t$ and reward $r_t$, the unitary-feedback model can be implemented by a controlled SU(2) rotation on the decision subspace, realizable by waveplates or an equivalent polarization controller. In the asymmetric case,
\begin{align}
U_{A,\mathrm{win}} &= R_{01}(-\theta), & U_{A,\mathrm{lose}} &= R_{01}(+\phi),\\
U_{B,\mathrm{win}} &= R_{01}(+\theta), & U_{B,\mathrm{lose}} &= R_{01}(-\phi),
\end{align}
with $\phi/\theta$ determined as in Supplementary Sec.~\ref{SI:secAppB}. In this alternative architecture, the auxiliary qutrit mode is not used by the ordinary two-choice feedback but remains available for the contextual probes.

\subsection*{KCBS probing measurements}
For the alternative finite-time learning experiment of Supplementary Sec.~\ref{SI:app:alt-rotation-reachability}, the target KCBS effects are the five fixed projectors $Q_j$ of Eq.~\eqref{SI:eq:new-fixed-probes}, not the auxiliary-axis $P_i$ illustrated in Supplementary Sec.~\ref{SI:secA1}. Adjacent $Q_j,Q_{j+1}$ are jointly measurable as the three-outcome measurement $\{Q_j,\,Q_{j+1},\,I-Q_j-Q_{j+1}\}$. The same calibrated $Q_j$ must be realized in either neighboring context.

Each $Q_j$ can be measured by a unitary $V_j$ mapping $|v_j\rangle$ to a reference mode (e.g., $|0\rangle$), followed by a click/no-click measurement on that mode:
\begin{equation}
\langle Q_j\rangle = \Pr(\text{click in }|0\rangle \text{ after }V_j).
\end{equation}
Suitable mode transformations could combine calibrated mixing between paths and polarization control. Individual binary implementations do not automatically verify joint compatibility; the required joint three-outcome tests and matched event effects must be calibrated independently.

\noindent\textit{Path-interference implementation of contextual probes.}
In the present encoding, contextual probe measurements rely on coherent interference between the modes
$|u,H\rangle$ and $|d,H\rangle$.
Operationally, the $H$-polarized components of the upper and lower paths are first separated using polarizing beam splitters.
These two modes are combined in a calibrated interferometer. A balanced beam splitter realizes equal-weight superpositions; general KCBS projectors require controllable mixing ratios as well as phases.
For example, balanced mixing followed by one-port detection projects onto $(|u,H\rangle+e^{i\delta}|d,H\rangle)/\sqrt2$, where $\delta$ is the controlled relative phase.
This measurement does not reveal which path the photon has taken, but instead probes the internal state in a
superposition basis, thereby implementing an incompatible measurement context.

The optical layout specifies possible mode transformations. A physical test would also require verified repeatable state preparation, alignment, compatibility, loss characterization, and an appropriate statistical analysis.
\section{Fixed-angle reward feedback in a weak-readout QTOW model}
\label{SI:app:alt-rotation-model}

\subsection{Why an incomplete decision readout is considered}
In a TOW bandit-decision system, successive choices and observed rewards alter the coupled preferences that guide later decisions. An \emph{incomplete readout} is one possible device design for making a preference-sensitive decision without identifying the internal preference perfectly. Neither noisy classical readout nor persistence of learning information implies quantum probability. As established in Sec.~\ref*{sec:classical-first}, an elementary classical two-state learner suffices for L1--L4, and the additional context-local diagnostic has its own calibration assumptions. This qutrit model provides a finite-time reward-conditioned KCBS construction using unitary feedback rather than the recovery--reset dynamics of the main manuscript.

Figure~\ref{SI:fig:qtow_concept} summarizes three successive operations: the binary decision instrument; the external bandit that supplies a Bernoulli reward; and the feedback operation conditioned on both the chosen action and that reward. Sharp projective KCBS probes are optional \emph{diagnostics} on prepared copies of the state and are not substituted for the binary decision instrument in normal learning trials.

\begin{figure}[t]
\centering
\resizebox{0.94\linewidth}{!}{%
\begin{tikzpicture}[
  font=\small,
  node distance=15mm and 18mm,
  >=Latex,
  box/.style={draw, rounded corners, line width=0.6pt, align=center, inner sep=6pt, minimum width=32mm},
  sbox/.style={draw, rounded corners, line width=0.6pt, align=center, inner sep=5pt, minimum width=28mm},
  arrow/.style={->, line width=0.8pt},
  darrow/.style={->, dashed, line width=0.7pt}
]
\node[box] (state) {Internal state\\\textbf{qutrit }$|\psi_t\rangle$};
\node[box, right=of state] (decision) {Decision instrument\\\textbf{$\{K_A,K_B\}$}\\outcome $d_t\in\{A,B\}$};
\node[box, right=of decision] (env) {Bandit environment\\\textbf{chosen arm $d_t$}\\reward probability $P_{d_t}$};
\node[box, below=of decision] (update) {Norm-preserving feedback\\\textbf{$U_{d_t,r_t}$}\\decision- and reward-conditioned};
\node[sbox, above=of decision] (aux) {Optional contextual probe\\\textbf{$Q_i$}\\uses qutrit auxiliary mode};

\draw[arrow] (state) -- (decision);
\draw[arrow] (decision) -- (env);
\draw[arrow] (env) -- (update);
\draw[arrow] (update) -- (state);
\draw[darrow] (state) -- (aux);
\draw[darrow] (aux) -- (decision);

\node[align=center, font=\footnotesize] at ($(update.south)+(0,-0.95)$) {%
This construction uses a qutrit with\\
two decision modes plus one auxiliary mode, with KCBS-compatible probes available.};
\end{tikzpicture}
}
\caption{Conceptual structure of QTOW. In the ordinary learning trajectory, the qutrit undergoes a two-outcome decision instrument $\{K_A,K_B\}$. For finite sharpness, the instrument perturbs the state while preserving part of the pre-decision state information. The selected option $d_t$ and reward $r_t$ then determine a norm-preserving feedback operation $U_{d_t,r_t}$. The third level is an auxiliary mode of the present construction and supports the KCBS-type projective contextuality diagnostic used below.}
\label{SI:fig:qtow_concept}
\end{figure}
\subsection{A complete decision--reward--feedback trial}
\label{SI:subsec:qtow-trial}
We work in the qutrit space $\mathcal H=\mathrm{span}\{|A\rangle,|B\rangle,|\perp\rangle\}$ and retain the following binary decision effects
\begin{align}
 E_A &=\operatorname{diag}\!\left(\frac{1+\eta}{2},\frac{1-\eta}{2},\frac12\right),&
 E_B&=I-E_A,& K_d&=\sqrt{E_d},\qquad 0<\eta<1.
 \label{SI:eq:new-native-effects}
\end{align}
The result $d\in\{A,B\}$ occurs with $p_d=\langle\psi|E_d|\psi\rangle$ and leaves the normalized post-decision state $|\psi^{(d)}\rangle=K_d|\psi\rangle/\sqrt{p_d}$. The choice $d$ is sent to an independent two-armed slot machine, which generates $r\in\{0,1\}$ with
\begin{equation}
 \Pr(r\mid d)=P_d^{\,r}(1-P_d)^{1-r}.\label{SI:eq:new-native-reward}
\end{equation}
The resulting state is updated \emph{after the reward} by a decision-and-reward-dependent operation $U_{d,r}$. For a branch in which the feedback angle is fixed in advance, the normalized state and the \emph{unconditional joint branch probability} are
\begin{align}
 |\psi'_{d,r}\rangle&=\frac{U_{d,r}K_d|\psi\rangle}{\sqrt{p_d}},\\
 \Pr(d,r\mid\psi)&=p_d P_d^{\,r}(1-P_d)^{1-r}. \label{SI:eq:new-native-branch}
\end{align}
The reward likelihood multiplies the \emph{probability of a branch}, not the conditional normalized state. The update depends on the \emph{observed} choice and reward, not a hypothetical expectation of the reward.

For states in the real decision plane, write $|\psi(q)\rangle=\cos q|A\rangle+\sin q|B\rangle$. The \emph{observed} A-choice probability is
\begin{equation}
 \Pr(A\mid q)=\frac{1+\eta\cos2q}{2},\label{SI:eq:new-native-choice}
\end{equation}
whereas $\cos^2q$ is an internal A-weight, \emph{not} the binary decision probability for $0<\eta<1$. After observing A and B, respectively, the angles obey
\begin{align}
 q_A&=\operatorname{atan2}\!\left(\sqrt{1-\eta}\sin q,\sqrt{1+\eta}\cos q\right),\\
 q_B&=\operatorname{atan2}\!\left(\sqrt{1+\eta}\sin q,\sqrt{1-\eta}\cos q\right).\label{SI:eq:new-native-post-choice}
\end{align}
One member of the model family uses symmetric feedback: $U_{A,1}=R_{AB}(-\theta)$, $U_{A,0}=R_{AB}(+\theta)$, $U_{B,1}=R_{AB}(+\theta)$, and $U_{B,0}=R_{AB}(-\theta)$ on the decision plane, acting as the identity on $|\perp\rangle$. These signs determine the direction of one update; they do not alone establish an asymptotic reward advantage. We define $R_{AB}(\beta)|\psi(q)\rangle=|\psi(q+\beta)\rangle$, equivalently $R_{AB}(\beta)=\left(\begin{smallmatrix}\cos\beta&-\sin\beta\\\sin\beta&\cos\beta\end{smallmatrix}\right)$ in the ordered basis $(|A\rangle,|B\rangle)$. Here $\theta>0$ is an amplitude angle (half the Bloch rotation angle).

For trajectories constrained to $q\in[0,\pi/2]$, a state-tracked boundary rule is
\begin{equation}
q'=\min\!\left\{\frac\pi2,\max\{0,q_d+s_{d,r}\theta\}\right\},\quad
s_{A,1}=s_{B,0}=-1,\quad s_{A,0}=s_{B,1}=+1.
\label{SI:eq:clip-policy}
\end{equation} This rule acts on a pure-state trajectory whose angle is available to the controller. A preparation-independent linear quantum channel on unknown mixed states instead uses the globally fixed rotation without clipping; both prescriptions agree for the interior first-trial calculation below.

\subsection{Optional recovery and operation reuse are not native requirements}
The main text uses the recovery-and-reset realization of Sec.~\ref*{sec:qtow-implementation}, which has useful stationary gain and an unconditional contextuality witness. The present alternative changes the reward update to amplitude rotations and demonstrates a distinct finite-horizon mechanism.

\section{Finite-horizon contextuality in the alternative fixed-angle feedback model}
\label{SI:app:alt-rotation-reachability}

\subsection{An admissible KCBS probe family, fixed before learning}
We use the standard sharp KCBS exclusive pentagon, but \emph{orient its symmetry axis along the learnable $|A\rangle$ direction}. The geometry is specified before the trial and remains fixed through the reward updates. An alternative orientation with its axis in the auxiliary direction is analyzed in Supplementary Sec.~\ref{SI:secA1}. Let $c^2=1/\sqrt5$, $s^2=1-1/\sqrt5$, and $\varphi_j=4\pi j/5$ for $j=0,\ldots,4$. Define
\begin{equation}
 |v_j\rangle=c|A\rangle+s\cos\varphi_j|B\rangle+s\sin\varphi_j|\perp\rangle,
 \qquad Q_j=|v_j\rangle\langle v_j|.\label{SI:eq:new-fixed-probes}
\end{equation}
Since $c^2+s^2\cos(4\pi/5)=0$, neighboring projectors are orthogonal, $Q_jQ_{j+1}=0$ (indices modulo five). Each neighboring pair admits one sharp three-outcome test $\{Q_j,Q_{j+1},I-Q_j-Q_{j+1}\}$, and the same $Q_j$ appears in both neighboring contexts. Set
\begin{equation}
 a:=\frac{5-\sqrt5}{2},\qquad \Sigma:=\sum_{j=0}^4Q_j
 =\operatorname{diag}(\sqrt5,a,a).\label{SI:eq:new-KCBS-sum}
\end{equation}
Thus on the ordinary learning plane
\begin{equation}
 K(q):=\langle\psi(q)|\Sigma|\psi(q)\rangle
 =a+(\sqrt5-a)\cos^2q.\label{SI:eq:new-KCBS-angle}
\end{equation}
An auxiliary-axis orientation, also valid geometrically, has $\Sigma_{\perp}=\operatorname{diag}(a,a,\sqrt5)$ and yields the constant $K_{\perp}=a\approx1.381966$ throughout the learning plane (Supplementary Sec.~\ref{SI:secA1}). The present fixed choice of probe geometry permits the learning state itself to cross the contextuality threshold.

\subsection{A positive-probability full learning cycle with a feedback-triggered violation}
\label{SI:subsec:qtow-one-step}
\noindent\textbf{Proposition (one-trial conditional reachability without boundary control).}
Consider Eqs.~\eqref{SI:eq:new-native-effects}--\eqref{SI:eq:new-native-branch} with initial $|\psi_0\rangle=(|A\rangle+|B\rangle)/\sqrt2$, $\eta=0.4$, $\theta=0.1$ rad, $P_A=0.8$ and $P_B=0.2$. Use the \emph{fixed} probes of Eq.~\eqref{SI:eq:new-fixed-probes}. Then the initial state does not violate KCBS; observing A does not yet produce a violation; but observing A and an external win, followed by the native feedback rotation, produces $K>2$ with positive probability $0.4$.

\emph{Proof.} For the initial state, $\Pr(A)=1/2$ by Eq.~\eqref{SI:eq:new-native-choice}. The A-branch after measurement has
\begin{equation}
 q_A=\arctan\sqrt{\frac{1-\eta}{1+\eta}}=0.5796397404\ldots,
 \qquad q_1=q_A-\theta=0.4796397404\ldots,\label{SI:eq:new-one-step-angles}
\end{equation}
and $0<q_1<q_A<\pi/2$, so the model's boundary correction is not activated. Evaluating Eq.~\eqref{SI:eq:new-KCBS-angle},
\begin{align}
 K(\psi_0)&=1.8090169944\ldots<2,\\
 K(\psi_0^{(A)})&=1.9798373876\ldots<2,\\
 K(\psi_1^{(A,\mathrm{win})})&=2.0541912687\ldots>2.\label{SI:eq:new-one-step-K}
\end{align}
The joint probability of that observed decision-and-reward history is $\Pr(A,\mathrm{win})=(1/2)P_A=0.4$. This is a \emph{real} reward-conditioning event, not a reward assigned by the observer after the fact. \hfill$\square$

Figure~\ref{SI:fig:new-learning-one-step} shows the threshold crossing. The four decision--reward branches and their unconditional one-trial average, $K=1.8556723440\ldots<2$, are reported in Table~\ref{SI:tab:one-trial-branches} of Supplementary Sec.~\ref{SI:app:full-learning-data}.

There is a \emph{non-fine-tuned} parameter interval for this feedback-triggered crossing. Define $q_c=\arccos\sqrt{(2-a)/(\sqrt5-a)}=0.553574\ldots$ rad (approximately $31.717^\circ$). For $\theta=0.1$, the choice-only A branch is nonviolating while the rewarded branch is violating provided
\begin{equation}
 \boxed{\ \cos[2(q_c+0.1)]<\eta\leq\frac1{\sqrt5}\ }
 \quad\text{i.e.,}\quad 0.26060385\ldots<\eta\leq0.44721360\ldots .
 \label{SI:eq:new-reward-crossing-region}
\end{equation}
No additional state-dependent feedback is needed for this first trial.

\begin{figure}[t]
 \centering
 \includegraphics[width=.85\linewidth]{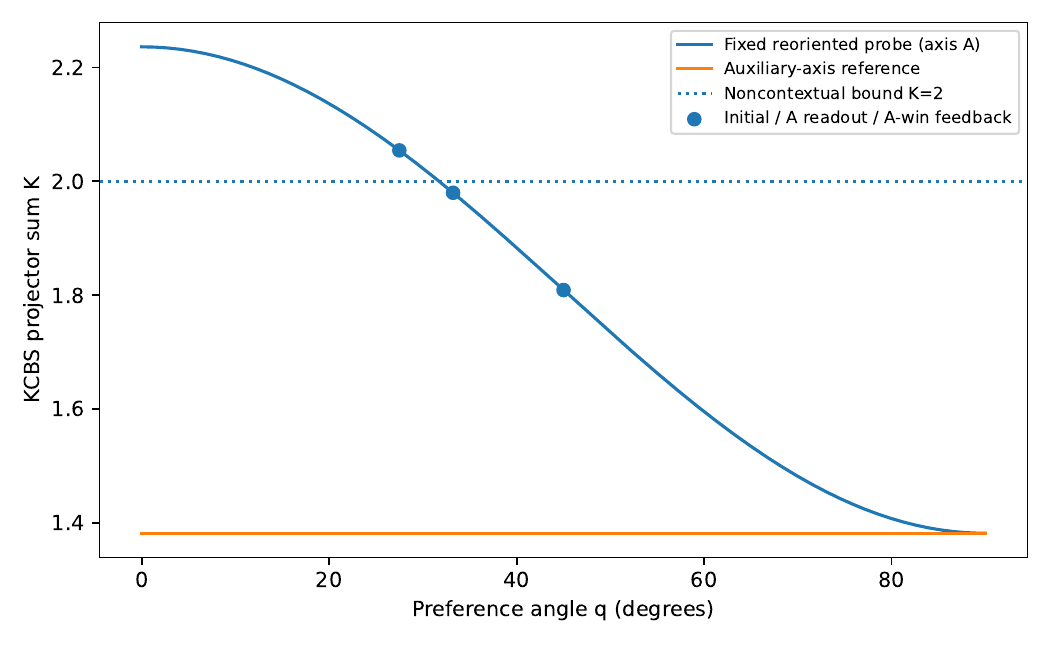}
 \caption{\textbf{Fixed-probe learning witness.} The auxiliary-axis pentagon does not violate on the ordinary learning subspace. The pentagon oriented along $|A\rangle$ has $K<2$ on the initial state and after the A readout, but $K>2$ following an A decision, external reward, and feedback. Angles refer to state amplitudes.}
 \label{SI:fig:new-learning-one-step}
\end{figure}

\subsection{Finite-trial statistics, conditional versus ensemble behavior}
\label{SI:subsec:qtow-multistep}
The single-trial result is independent of any boundary convention. For finite-trial predictions, we computed all $4^n$ decision/reward histories for $n\leq8$ under \emph{two explicitly different feedback specifications}: (C) pure-state, history-conditioned angle clipping to $[0,\pi/2]$, and (U) the same fixed feedback rotation on the full real decision plane without clipping. Specification (U) defines a state-independent unitary feedback channel and allows signed real amplitudes; specification (C) tracks pure-state angles within the nonnegative-amplitude chart. They agree in the first complete trial.
The finite-horizon statistics are summarized in Table~\ref{SI:tab:ensemble-trials} of Supplementary Sec.~\ref{SI:app:full-learning-data}, with additional matrix checks in Supplementary Sec.~\ref{SI:app:new-learning}. The (U) model, unlike the state-tracked pure-trajectory protocol (C), defines a single channel on arbitrary density matrices.

For the state-independent fixed-rotation specification (U), there is also a short \emph{analytic} check. Let $x=\langle\sigma_x\rangle$ and $z=\langle\sigma_z\rangle$ be the two real Bloch components of the unconditional qubit substate and put $\ell=\sqrt{1-\eta^2}$, $\Delta=P_A-P_B$ and $g=P_A+P_B$. One feedback trial with fixed $\theta$ acts linearly as
\begin{equation}
 \begin{pmatrix}x'\\z'\end{pmatrix}=
 \begin{pmatrix}\ell\cos2\theta&-\Delta\sin2\theta\\
 \ell\Delta\sin2\theta&\cos2\theta\end{pmatrix}
 \begin{pmatrix}x\\z\end{pmatrix}
 +\begin{pmatrix}\eta(1-g)\sin2\theta\\0\end{pmatrix}.
 \label{SI:eq:new-bloch-fixed}
\end{equation}
For the above parameters, $g=1$ and $(x_0,z_0)=(1,0)$. Equation~\eqref{SI:eq:new-bloch-fixed} gives $z_6=0.458457111\ldots$ and
\begin{equation}
 \mathbb E_{\rm U}[K_6]=a+(\sqrt5-a)\frac{1+z_6}{2}
 =2.0048015545\ldots>2,\label{SI:eq:new-sixth-ensemble}
\end{equation}
independently matching explicit enumeration. Thus an \emph{unconditional, finite-trial ensemble KCBS violation} is available in this well-defined state-independent extension. Figure~\ref{SI:fig:new-learning-ensemble} displays the finite-horizon ensemble values. In the contractive regime at $g=1$, the fixed-angle model approaches a neutral asymptotic state, even though it shows finite-time reward responsiveness; its stationary performance therefore differs from that of the classical learner in Sec.~\ref*{sec:classical-first}.

\begin{figure}[t]
 \centering
 \includegraphics[width=.82\linewidth]{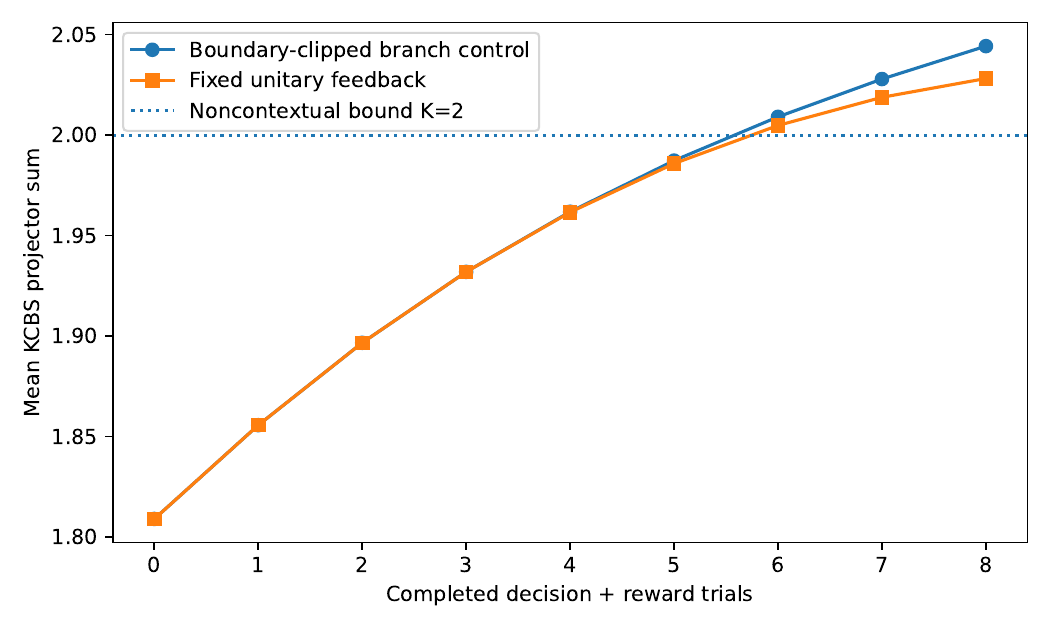}
 \caption{\textbf{Unconditional learning dynamics for eight finite trials.} Weighted sums over all $4^n$ histories; the pure-state clipping convention and the globally fixed unitary feedback are shown separately. Both have $K>2$ in the finite-time ensemble by trial six. The fixed-angle version is a linear quantum channel; the clipped version is a controlled pure-state trajectory.}
 \label{SI:fig:new-learning-ensemble}
\end{figure}

\subsection{Operational meaning of the fixed-probe test}
In the sharp KCBS test, five binary events are operationally shared across their neighboring contexts, and adjacent Yes events are exclusive. Under those premises, $K>2$ precludes one context-independent deterministic assignment of all five outcomes, although it does \emph{not} preclude a classical computer from sampling context-labelled outcomes or maintaining a continuous ray-state simulator executing the specified measurement/update formulas. The standard KCBS diagnostic is a test on the \emph{separate, sharp} projectors $Q_j$, not on the two ordinary unsharp A/B decision effects $E_d$.

The corresponding repeated-preparation test and its compatibility checks are specified in Sec.~\ref*{sec:operational-comparison}. The experiment must distinguish the fixed, sharp diagnostic from the unsharp learning readout.

\subsection{Connection with operational minimality}
The preceding sections distinguish three claims. First, classical TOW reward learning and even shared-event context-local sampling admit fully classical stochastic realizations. Second, a \emph{noncontextual completion} of a chosen shared-event diagnostic may require a nonzero semantic-repair cost, and independent composition/calibration constraints can restrict the accessible KCBS values; neither claim assumes a Born probability rule. Third, a unified qutrit QTOW implementation, with an explicitly fixed reoriented diagnostic family, makes KCBS contextuality accessible through its \emph{actual} reward-conditioning dynamics. The qutrit construction provides a physical representation of the extended diagnostic service; it shows explicitly how a reward-conditioned learning state and a compatible family of contextual tests can operate in the same device.
\section{Technical checks for the fixed-probe learning witness}
\label{SI:app:new-learning}
\subsection*{Pentagon normalization and projector sum}
Writing $c^2=1/\sqrt5$ and $s^2=1-1/\sqrt5$, neighboring KCBS vectors satisfy
$\langle v_j|v_{j+1}\rangle=c^2+s^2\cos(4\pi/5)=0$.
Using $\sum_j\cos\varphi_j=\sum_j\sin\varphi_j=\sum_j\sin\varphi_j\cos\varphi_j=0$ and
$\sum_j\cos^2\varphi_j=\sum_j\sin^2\varphi_j=5/2$ gives
$\Sigma=\operatorname{diag}(5c^2,5s^2/2,5s^2/2)$.
In particular, the fixed geometry is one family chosen \emph{before} either the experiment or the reward outcome. The underlying $\{Q_j\}$ may be interrogated in five separately repeated neighboring contexts.

\subsection*{Verification of the reward-triggered parameter interval}
Let $u_c=(2-a)/(\sqrt5-a)$ and $q_c=\arccos\sqrt{u_c}$. For the neutral input, the A-readout angle satisfies
$\cos^2q_A=(1+\eta)/2$ and $q_A\in(0,\pi/4)$. Thus the state after choice A remains below the classical threshold if $\eta\le2u_c-1=1/\sqrt5$. The A-win state exceeds the threshold if $q_A-\theta<q_c$, equivalently $\eta>\cos 2(q_c+\theta)$, provided $0<\theta<q_A$ so that the boundary is not reached. At $\theta=0.1$, this gives the nonempty interval in Eq.~\eqref{SI:eq:new-reward-crossing-region}. The external Bernoulli reward enters only through the joint branch weight $P_A/2$ for the first neutral trial; it is essential for selecting which update is applied.

\subsection*{Finite-trial reproduction and model scope}
The reproducibility script constructs the three-dimensional $K_A,K_B$ and $U_{d,r}$ matrices and compares the four branch angles and KCBS values with direct normalized matrix multiplication. It also exhaustively propagates all branch paths through five trials as a matrix-vs-angle check (1,364 propagated branches in total), and enumerates all $4^n$ branch histories through eight trials for two explicitly different boundary policies. The fixed-angle unconditional result is separately checked by Eq.~\eqref{SI:eq:new-bloch-fixed}; the probability weights sum to one at every depth. The data files report all four first-trial histories and the ensemble statistics, avoiding a conditional/unconditional ambiguity. The state-dependent clipping convention requires an explicit state/history-tracking control if it is to be implemented on unknown mixtures, and is not asserted to be a unique state-independent CPTP channel.
\section{Detailed branch and finite-trial learning calculations}\label{SI:app:full-learning-data}
The four observed branches, computed by the same Kraus and feedback maps, are shown in Table~\ref{SI:tab:one-trial-branches}.
\begin{table}[ht]
\centering\small
\caption{Complete decision--reward branches for the neutral initial state.}
\label{SI:tab:one-trial-branches}
\begin{tabular}{lrr}
\hline
History after one complete trial & Probability & $K$ after feedback\\
\hline
$A$, win & 0.400 & 2.0541913\\
$A$, loss & 0.100 & 1.8986734\\
$B$, win & 0.100 & 1.5638427\\
$B$, loss & 0.400 & 1.7193606\\
\hline
\end{tabular}
\end{table}
The corresponding one-trial \emph{unconditional} ensemble is not KCBS-violating:
\begin{equation}
 \operatorname{Tr}(\rho_1\Sigma)=\sum_{d,r}\Pr(d,r)K(\psi_1^{(d,r)})=1.8556723440\ldots<2.\label{SI:eq:new-one-step-mean}
\end{equation}
The conditional result is therefore not presented as an ensemble-averaged one-step violation. The KCBS statistic must be estimated by re-preparing the \emph{same} conditioned state on many independently run trials and distributing those copies among the five fixed neighboring contexts.

\subsection*{All branch and ensemble values for the two update conventions}
Table~\ref{SI:tab:ensemble-trials} compares the two controls at selected finite horizons.
\begin{table}[ht]
\centering\small
\caption{Finite-time probability of a KCBS violation and the unconditional KCBS sum for the clipped (C) and fixed-angle (U) protocols.}
\label{SI:tab:ensemble-trials}
\begin{tabular}{rrrrr}
\hline
 & \multicolumn{2}{c}{C: history-controlled clipping} & \multicolumn{2}{c}{U: fixed rotation}\\
$n$ & $\Pr(K>2)$ & $\mathbb E[K]$ & $\Pr(K>2)$ & $\mathbb E[K]$\\
\hline
0 & 0 & 1.809017 & 0 & 1.809017\\
1 & 0.400000 & 1.855672 & 0.400000 & 1.855672\\
2 & 0.289288 & 1.896650 & 0.289288 & 1.896650\\
4 & 0.540011 & 1.961852 & 0.540011 & 1.961616\\
6 & 0.641963 & 2.009084 & 0.641963 & 2.004802\\
8 & 0.709045 & 2.044256 & 0.699063 & 2.028154\\
\hline
\end{tabular}
\end{table}

\subsection*{Native protocol has an exact classical trajectory simulator}
For an initial known real pure state $q_0$, a classical real-valued state $q_t$ reproduces all native action/reward outcome strings: sample $d_t$ with $\Pr(A\mid q_t)=(1+\eta\cos2q_t)/2$; sample $r_t\sim\mathrm{Bernoulli}(P_{d_t})$; compute $q_{d_t}$ using Eq.~\eqref{SI:eq:new-native-post-choice}; and apply the declared outcome-dependent feedback to obtain $q_{t+1}$. Conditional probabilities agree by construction at every step, and multiplication gives equality of every finite sequence probability. This is an exact probabilistic construction, not an empirical approximation. It says nothing about noncontextual global assignments for a richer family of sharp diagnostic events. For general mixed input preparations, the pure-angle state alone is insufficient unless the preparation decomposition is provided; the corresponding fixed-angle linear average model can use the two real Bloch variables of Eq.~\eqref{SI:eq:new-bloch-fixed}.

\subsection*{Why pure-state boundary clipping is not a single mixed-state channel}
The pure-state boundary prescription that chooses a reduced rotation angle after accessing the \emph{pure} post-measurement angle $q$ defines a stochastic controller on individually tracked pure trajectories. It need not induce an affine map on an arbitrary density operator $\rho$. An explicit comparison of distinct decompositions of the same real two-dimensional density matrix found different probabilities for a later three-symbol event of the form $(A,\mathrm{win},A)$ under componentwise clipping. Thus an unspecified mixed input cannot be evolved by one fixed CPTP map implementing this prescription. It is consistent to use a classical register carrying the conditioned pure angle, or to study the globally fixed-angle unitary model as a different, well-defined instrument. Neither alternative is silently identified with the other, and the interior one-trial witness is unaffected.
\section{Historical and physical motivation for TOW}\label{SI:app:tow-background}
\noindent This appendix places the TOW bandit-decision method and its reward-based adaptation in the context of physical and natural computation.
\subsection*{From physics-agnostic computation to physics-based decision making}

Much of digital computation uses an abstract logical description intended to be largely independent of the details of a given hardware implementation \cite{SI:Herken1995}.
CMOS implementations use physical design and circuit techniques to control noise and sustain reliable logical operation \cite{SI:Sze1981}.
The resulting design trade-offs can involve energy, reliability and implementation complexity \cite{SI:Baker2010}.

In contrast, information processing in natural systems is intrinsically tied to physical constraints.
Biological and physical decision-making processes operate under conservation laws, continuity, and stochastic fluctuations, and often exploit---rather than eliminate---these features \cite{SI:Castro,SI:Kari,SI:Aono}.
This observation motivates a \emph{physics-based} approach to decision making, in which adaptive behavior is implemented directly through constrained physical dynamics rather than abstract logical rules imposed from above.

This contrast is not intended as a strict identification of digital computation with artificial intelligence or of physics-based computation with natural intelligence. Rather, it provides a controlled setting for comparing two ways of realizing adaptive decision processes: by explicitly representing computational states and contextual information, or by allowing part of the decision structure to be carried intrinsically by the dynamics of the underlying system. In this sense, TOW provides a bridge between natural, physics-grounded decision mechanisms and resource-constrained artificial decision systems.

\subsection*{Tug-of-War principle as a physics-grounded decision mechanism}

The Tug-of-War (TOW) method was proposed as a \emph{physics-inspired decision-making} mechanism: resource conservation couples the option preferences, while the reward-dependent update law is designed to exploit that coupling \cite{SI:TOW2009,SI:TOW2010,SI:TOWBio}.
It provides a reward-adaptive decision policy for multi-option bandit problems, a standard setting for reinforcement learning, in which an agent selects among alternatives and receives stochastic rewards \cite{SI:TOW2015,SI:TOWB,SI:TOWAS}.

In the classical TOW model, the decision maker maintains internal variables associated with each option, representing preference scores coupled by a resource constraint; these scores need not be unbiased estimates of arm utilities.
At each step, a choice is made by comparing these internal quantities, biasing selection toward the option with the larger value. Crucially, these variables are not updated independently.

The defining feature of the TOW mechanism is a conservation constraint:
an increase in preference for one option necessarily entails a compensating decrease for the other.
This coupling reflects the reallocation of a finite internal resource, rather than the independent accumulation of evidence.
As a result, competition between alternatives is embedded directly into the internal dynamics of the decision process.

Stochastic fluctuations can also support exploration in TOW implementations.
Random perturbations can sustain exploration and reduce the risk of remaining committed to an unfavorable option; the resulting decision quality depends on the environment and update parameters.
Their interaction with the resource-coupled update can help manage exploration and exploitation, depending on the decision rule and the reward environment.

\subsection*{Decision Maker: TOW devices and applications}

This physics-grounded principle has been implemented in a wide range of physical substrates, including quantum dots \cite{SI:Qdot,SI:QdotE}, photonic systems \cite{SI:Single,SI:Single2}, atomic switches \cite{SI:TOWAS}, semiconductor lasers \cite{SI:Ultra}, ionic devices \cite{SI:Ion}, and resistive memories \cite{SI:Resist}.
These studies illustrate possible physical implementations of resource-coupled bandit decisions, with varying levels of external control and feedback.

The lightweight structure of many TOW updates also motivates implementations on resource-constrained hardware.
Such simplicity may also favor embedding reward-adaptive decision policies in existing hardware.
Indeed, TOW-based mechanisms have been applied to practical problems such as autonomous channel selection, fair resource allocation, and adaptive control in Internet-of-Things (IoT) communication systems, where decision making must be performed locally, efficiently, and with minimal computational overhead \cite{SI:Comm1,SI:Comm2,SI:CommK,SI:Fair}.

Taken together, these realizations demonstrate that the TOW mechanism is not tied to any specific physical platform or computing architecture, but instead reflects a general decision-making principle grounded in physical constraints and exploitable across a broad range of technological settings.

\endgroup
\end{document}